\documentclass[opre,nonblindrev]{informs3modified} % current default for manuscript submission
\usepackage[english,american]{babel}

\usepackage{enumerate}
\usepackage{epstopdf}
\usepackage[pdftex,colorlinks=true,urlcolor=blue,citecolor=black,anchorcolor=black,linkcolor=black]{hyperref}
\DoubleSpacedXI % Made default 4/4/2014 at request
\usepackage{natbib}
 \bibpunct[, ]{(}{)}{,}{a}{}{,}%
 \def\bibfont{\small}%
\TheoremsNumberedThrough     % Preferred (Theorem 1, Lemma 1, Theorem 2)
\newtheorem{defination}{Definition}
\makeatletter
\providecommand{\@LN}[2]{}
\makeatother
\ECRepeatTheorems

\EquationsNumberedThrough    % Default: (1), (2), ...
\begin{document}
%%%%%%%%%%%%%%%%

%\setcounter{page}{1}
%\DOI{10.1287/xxxx.0000.0000}%

\RUNAUTHOR{Plumlee and Lam}

\RUNTITLE{An Uncertainty Quantification Method for Inexact Simulation Models}

% \TITLE{Bundling Information Goods of Decreasing Value}
\TITLE{An Uncertainty Quantification Method  for Inexact Simulation Models}

\ARTICLEAUTHORS{%
\AUTHOR{Matthew Plumlee, Henry Lam}
\AFF{Department of Industrial and Operations Engineering, University of Michigan, Ann Arbor, MI, USA}
} % end of the block
\ARTICLEAUTHORS{%
\AUTHOR{Matthew Plumlee, Henry Lam}
\AFF{Department of Industrial and Operations Engineering, University of Michigan, Ann Arbor, MI, USA}
} % end of the block

\ABSTRACT{%
The vast majority of stochastic simulation models are imperfect in that they fail to exactly emulate real system dynamics. The inexactness of the simulation model, or model discrepancy, can impact the predictive accuracy and usefulness of the simulation for decision-making. This paper proposes a systematic framework to integrate data from both the simulation responses and the real system responses to learn this discrepancy and quantify the resulting uncertainty. Our framework addresses the theoretical and computational requirements for stochastic estimation in a Bayesian setting. It involves an optimization-based procedure to compute confidence bounds on the target outputs that elicit desirable large-sample statistical properties. We illustrate the practical value of our framework with a call center example and a manufacturing line case study.
}%
% Despite this, these imperfect models are still useful in practice, so long as one knows how the model is inexact.
% This inexactness is measured by a discrepancy between the proposed stochastic model and a true stochastic distribution across multiple values of some decision variables.

% \ABSTRACT{%
% The vast majority of stochastic simulation models are imperfect in that they fail to fully emulate the entirety of real dynamics.  Despite this, these imperfect models are still useful in practice, so long as one knows how the model is inexact.  This inexactness is measured by a discrepancy between the proposed stochastic model and a true stochastic distribution across multiple values of some decision variables. In this paper, we propose a method to learn this discrepancy using data collected from the system of interest. Our approach is a novel Bayesian framework that addresses both the theoretical and the computational requirements for estimation of probability measures. We investigate the large-sample statistical properties and illustrate the practical value of this approach with a running example.
% }%

\KEYWORDS{model calibration; simulation; Bayesian methods; optimization}
%
%\HISTORY{This paper was
%first submitted on April (), ()()()() and has been with the authors for
%()() years for ()() revisions.}

\maketitle
%%%%%%%%%%%%%%%%%%%%%%%%%%%%%%%%%%%%%%%%%%%%%%%%%%%%%%%%%%%%%%%%%%%%%%
%%
\section{Introduction} \label{sec:intro}
Stochastic or simulation models are only approximations to the reality. A conjectured model may not align with the true system because of unobserved complexity. Moreover, some highly accurate models, even if formulable, may not be implementable due to computational barriers and time constraints, in which case a simpler, lower-fidelity model is adopted. In all these cases, there is a discrepancy between the model and the reality, which we call \emph{model discrepancy}. This article describes a data-processing framework to integrate data from both a simulated response and the real system of interest, under the presence of model discrepancy, to reliably predict stochastic outputs of interest.

Our objective is motivated from everyday practice of simulation analysis. For example, this article describes  a major manufacturer that is interested in assessing the impact of the staffing level of support workers on a production line via discrete-event simulation.  Twelve weeks were spent carefully designing and tuning the simulation model and the final report included seventy-five realizations of the simulation model at each potential staffing level.   The limited amount of realizations gives rise to a simulation error (also termed a Monte Carlo error).  In addition, when data at the current staffing level was compared to the simulation model realizations, it was clear the simulation model was inaccurate.  Yet, given the resources already invested, the manufacturer was interested if the simulation model could still be used to guide the staffing level decision.  An approach that can account for both sources of errors can save significant costs and improve the decisions in situations like these.
% Situations like this occur frequently in e will demonstrate in this paper how to systematically correct and use such an inexact simulation model.

Differences between a simulation and real data is traditionally addressed during the important practice of \emph{model validation} and \emph{calibration} in the simulation literature, which refers to the joint task of checking whether a developed stochastic model sufficiently reflects the reality (validation), and if not, re-developing the model until it matches (calibration) (e.g., \cite{Sargent2013}, \cite{banks2000dm} Chapter 10, \cite{kelton2000simulation} Chapter 5). Conventional validation methods compare relevant outputs from simulation models and real-world data via statistical or Turing tests (e.g. \cite{schruben1980establishing} and \cite{balci1982some}). In the case of a mismatch, guided expert opinions, together with possibly more data collection, are used to re-calibrate the model recursively until acceptable accuracy \citep{sargent1998verification}.   While these tools are fundamentally critical to the practice of simulation, there can be two deficiencies when using calibration in an ad-hoc way:
\begin{enumerate}
\item  It necessitates building increasingly sophisticated models after unsatisfactory conclusions.  This process potentially places a heavy burden on a simulation modeler/software, consumes time and, moreover, may end up in non-convergence to an acceptable ultimate model.
\item The recursive refinement of the model to align it with the real data along the development process involves hidden parameter choices and simultaneous estimations. These details, which are often overlooked and unaccounted for, complicate statistically justified uncertainty quantification alongside prediction.
% \item  because there are hidden degrees of freedom that are unaccounted for in typical statistical predictions.
\end{enumerate}

Our goal is thus to investigate a framework that systematically offers predictive bounds using a simulation model without the traditionally encountered recursive efforts. Our framework is a stochastic version of model calibration that is similar in spirit to deterministic model calibration \citep{kennedy2001bayesian}. The basic idea is to view potential model discrepancy as an object that can be inferred statistically, or plainly put, to ``model" this potential error.  To conduct feasible inference, often the model discrepancy is assumed to have some structure decided a priori of observing data, and data are used to update the uncertainty on predictions of the true system. Since \cite{kennedy2001bayesian}, this idea has been extended and widely applied in various scientific areas, e.g., \cite{tuo2015efficient,higdon2004combining,plumlee2016bayesian}.  In the stochastic simulation literature, similar machinery has appeared under the heading of  stochastic kriging (\cite{ankenman2010stochastic,Staum2009,chen2013enhancing,chen2012effects,chen2014stochastic,chen2016efficient}). In the stochastic kriging literature, the oracle benchmark is the simulation model and stochastic kriging is used to reduce simulation effort by borrowing information from the simulation outputs at a collection of design values. In the model discrepancy setting, the oracle benchmark is the real system's probabilistic generating mechanism and our goal is to improve the prediction accuracy and quantification of uncertainties associated with the simulation model.

One challenge in bringing the deterministic model discrepancy machinery to stochastic simulation is that in the latter case, the inference objects are themselves embedded in probability spaces. The stochastic simulation model and the real system are naturally represented as probability distributions (think of the output distributions of a queueing or a stochastic inventory model), which constitute the basis of calculation in many decision-making tasks (for example, computing the chance that the outcome is in some region that indicates poor performance). Consequently, the learning and the uncertainty quantification of the discrepancies need to take into account the resulting probabilistic constraints. This is beyond the scope of the established inference and computation tools in the deterministic model discrepancy literature.

As our main contribution, we develop a framework to infer stochastic model discrepancies that is statistically justified and computationally tractable under the constraints discussed above. On the statistical aspect, we build a Bayesian learning framework that operates on the space of likelihood ratios as the representation of model discrepancies between simulation and reality. We study how this representation satisfies the constraints necessarily imposed in capturing stochastic model discrepancies and leads to desirable asymptotic behavior. On the computational aspect, we propose an optimization approach to obtain prediction bounds. Though sampling techniques such as Markov chain Monte Carlo \cite[Chapters 11 and 12]{gelman2014bayesian} are widely used in Bayesian computation, they encounter difficulties in our setting due to the constraints and high-dimensionality. Our approach, inspired from the recent literature in robust optimization (\cite{ben2002robust,ben2009robust,bertsimas2011theory}), alleviates this issue via the imposition of suitable optimization formulations over posterior high probability regions. We study the statistical properties of these formulations and demonstrate that they are equally tight in terms of asymptotic guarantees to traditional Bayesian inference.

We close this introduction by briefly reviewing two other lines of related work. First, in stochastic simulation, the majority of work in handling model uncertainty focuses on input uncertainty; see, e.g. the surveys \cite{barton2002panel,henderson2003input,chick2006bayesian,barton2012tutorial,song2014advanced,lam2016advancedtutorial}, \cite{nelson2013foundations} Chapter 7. They quantify the impacts on simulation outputs due to the statistical uncertainty in specifying the input models (distributions, stochastic assumptions etc.), assuming input data are available. Approaches include the delta method (\cite{cheng1997sensitivity}) and its variants such as the two-point method (\cite{cheng1998two,cheng2004calculation}), the bootstrap (\cite{barton1993uniform,barton2001resampling,cheng1997sensitivity}) which can be assisted with stochastic kriging-based meta-models (\cite{barton2013quantifying,xie2014bayesian}), and Bayesian methods (\cite{chick2001input,zouaoui2003accounting,zouaoui2004accounting,xie2014bayesian,biller2011accounting}). Added to these approaches are recent perspectives of model risks and robust optimization that do not necessarily directly utilize data (\cite{glasserman2014robust,lam2013robust,lam2011sensitivity,ghosh2015computing}). The second line of related work is queueing inference that investigates the calibration of input processes and system performances from partially observed queueing outputs such as congestion or transaction data (e.g., the queue inference engine; \cite{larson1990queue}). This literature utilizes specific queueing structures that can be approximated either analytically or via diffusion limits, and as such allow tractable inference. Techniques include maximum likelihood estimation (\cite{basawa1996maximum,pickands1997estimation}), nonparametric approaches (\cite{bingham1999non,hall2004nonparametric}) and point processes (\cite{whitt1981approximating}).
Recently, \cite{goeva2014reconstructing} study calibration of input distributions under more general simulation models. Like the input uncertainty literature, however, these studies assume correctly specified system logics that imply perfect matches of the simulation models with real-world outputs.

% All these literature assume correctly specified system logics and data collection at the input level.

\section{Stochastic Model Discrepancy: Setting and Notations} \label{sec:setting}
This section describes our setting and notations throughout this paper. We consider a system of interest that outputs a discrete random response over the space $\mathcal{Y}$ with cardinality $m$.  For notational simplicity, we will use the space $\mathcal Y=\{1,\ldots,m\}$.  This response depends on a vector of design variables, denoted $x$, which can be broadly defined to include input variables that are not necessarily controllable. We presume a finite set of design points or design values $x_j$, $j = 1,\ldots,s$.  The probability mass function $\pi_j=\{\pi_j(i)\}_{i=1,\ldots,m}$ describes the distribution of the response of the real system on $\mathcal Y$ under $x_j$.   Examples of the response include the waiting times in call centers \citep{brown2005statistical} and hospitals \citep{helm2014design}.    In the first example, design variables could be the number of servers, the system capacity, and the arrival rate.  In  the second example, the design variable could be the rate of elective admissions.
% The term design point will be understood to mean the point in the space corresponding to a combination of design variables.

The objective is to draw conclusions about $\pi_j$ for several $j$'s. These distributions form the basis in evaluating quantities of interest used for decision-making.  When responses are independently observed from the real system (e.g., from a designed experiment \citep{li2015value}), $n_j(i)/n_j$ is a reasonable estimate of $\pi_j$, where $n_j(i)$ counts the number of outcomes equal to $i$ and $n_j$ is the total number of recorded responses at $x_j$.  In the setting of simulation modeling, however, these empirical estimates are often inadequate because typical decision-making tasks, like feasibility or sensitivity tests, are applied on system configurations that are sparsely sampled or even never observed. This means that accurate empirical estimates for the $x_j$ values of interest are not available. In fact, for these $j$'s, $n_j$ can often times be $0$.

In contrast, using state-of-the-art understanding of the system, possibly simplified for computational concerns, an operations researcher builds a simulation model (typically based on discrete-event simulation) to estimate $\tilde{\pi}_{j}=\{\tilde\pi_j(i)\}_{i=1,\ldots,m}$, the simulated distribution of the response at the design point $x_j$. In parallel to the real responses, we denote $\tilde{n}_j(i)$ as the count of outcome $i$ and $\tilde n_j$ as the total number of replications in a simulation experiment at $x_j$, and $\tilde{n}_j(i)/\tilde{n}_j$ is hence an estimate of $\tilde\pi_j(i)$. Unlike the real responses, it is often affordable to generate a more abundant number of $\tilde n_j$ and hence a more accurate estimate of $\tilde\pi_j$.  However, the difference between $\tilde{n}_j(i)/\tilde{n}_j$ and $\tilde\pi_j(i)$ remains a source of uncertainty.

Our premise is that the real response distribution $\pi_j$ and the simulated distribution $\tilde\pi_j$ differ. Thus, in order to make conclusions about $\pi_j$, we must conjecture about the potential gap between $\pi_j$ and $\tilde\pi_j$ with the limited simulation and real-world data. The remainder of this section describes our framework for defining the discrepancy between $\pi_j$ and $\tilde\pi_j$.

First note that both $\pi_j$ and $\tilde\pi_j$ obviously must satisfy the criteria of a probability distribution:
\begin{defination} \label{def:valid_distribution}
Any mapping $p: \mathcal{Y} \rightarrow \mathbb{R}$ is a \emph{valid distribution} if
\begin{enumerate}[(i)]
\item $p(i)\geq0$ for all $i = 1,\ldots,m$ and
\item $\sum_{i=1}^m   p(i) =  1$.
\end{enumerate}
\end{defination}

We define the discrepancy between $\pi_j$ and $\tilde\pi_j$ as $\delta_j=\{\delta_j(i)\}_{i=1,\ldots,m}$  where
\begin{equation}
\delta_j(i)  =  \frac{\pi_j(i)}{\tilde{\pi}_j(i)}.  \label{def discrepancy}
\end{equation}
In other words, $\delta_j$ reflects the the ratio between the probabilities of the true responses and simulated responses. If $\delta_j (i) = 1$ for all $i$, the simulation model is correctly specified. Definition \ref{def discrepancy} is analogous to that of likelihood ratio in the context of importance sampling (e.g., \cite{mcbook}, Chapter 9; \cite{asmussen2007stochastic}, Chapter IV; \cite{glasserman2003monte}, Chapter 4). In the model risk literature, similar object as Definition \ref{def discrepancy} also appears as a decision variable in worst-case optimization problems used to bound performance measures subject to the uncertainty on the true model relative to a conjectured stochastic model (often known as the baseline model).   Examples include Gaussian models with mean and covariance uncertainty represented by linear matrix inequalities (\cite{hu2012robust}), and nonparametric uncertainty measured by Kullback-Leibler divergence (e.g., \cite{glasserman2014robust,lam2013robust}). Our definition \ref{def discrepancy} is along a similar vein as these work, but  rather than using it as a tool to speed up simulation (in importance sampling) or an optimization decision variable (in model risk), our $\delta_j$ is an object to be \emph{inferred} from data.

Note that \eqref{def discrepancy} is not the only way to define stochastic model discrepancy. Another natural choice, which more closely mimics the established deterministic counterpart \citep{kennedy2001bayesian}, is via
\[\pi_j(i)-\tilde{\pi}_j(i)\]
The choice of which version of discrepancy to use relates to the convenience in statistical modeling. We adopt the multiplicative version in \eqref{def discrepancy} based on its analog with likelihood ratio, which facilitates our inference.

Since $\pi_j$ and $\tilde\pi_j$ are valid distributions, the model discrepancy $\delta_j$ defined in \eqref{def discrepancy} must satisfy the following criteria with respect to $\tilde\pi_j$:
\begin{defination} \label{def:valid}
Say $p$ is a valid distribution with $p(i)> 0$.  $d$ is a \emph{valid discrepancy} with respect to $p$ if
\begin{enumerate}[(i)]
\item $d(i)\geq0$ for all $i = 1,\ldots,m$ and
\item $\sum_{i=1}^m   d(i) p(i)  =  1$.
\end{enumerate}
\end{defination}
Clearly, if $d$ is a valid discrepancy and $p$ is a valid distribution then $\{d(i) p(i)\}_{i=1,\ldots,m}$ will also be a valid distribution.

Definition \ref{def:valid} plays a vital role in our subsequent analysis as they characterize the properties of our inference targets. Unlike deterministic model discrepancies, these conditions come from the probabilistic structure that arises uniquely in stochastic model discrepancies. Note that Definition \ref{def:valid} coincides with that of a likelihood ratio (e.g., \cite{asmussen2007stochastic}).

Lastly, in addition to model discrepancy, simulation noise and experimental noise also contribute to the uncertainty in estimating $\pi_i$, i.e., the noise of the estimator $n_j(i)/n_j$ for $\pi_j(i)$ and $\tilde n_j(i)/\tilde n_j$ for $\tilde\pi_j(i)$. Our analysis will also incorporate these sources of uncertainty.
\section{A Bayesian Framework} \label{sec:learn}
We propose a Bayesian framework to infer the discrepancy $\delta_j$. The framework has the capability to quantify uncertainty under limited data environments (common in our setting where observed responses from the real system may be sparse or absent for some design points), and to incorporate prior information that  anticipates similar discrepancies for similar design points, where the similarity is measured by the distance between the design values. We will also see how the framework can account for the notion of a valid discrepancy provided in Definition \ref{def:valid}.

The term \emph{data} substitutes for the collection of all observed responses from the real system and the simulation model, which is sufficiently represented as
\[\text{data} = \left\{n_j(i),  i = 1,\ldots,m, j = 1,\ldots, s  \text{ and } \tilde{n}_j(i),  i = 1,\ldots,m,  j = 1,\ldots, s \right\}.\]
Our main inference procedure is the Bayes rule summarized as
\begin{equation}
\operatorname{post}\left(d,\tilde{p},\text{data}\right) \propto \operatorname{likelihood}(d,\tilde{p},\text{data} ) \operatorname{prior}(d,\tilde{p}), \label{Bayesian update}
\end{equation}
where $d$ and $\tilde{p}$ are the locations at which the density is evaluated for $\delta=(\delta_j)_{j=1,\ldots,s}$ and $\tilde{\pi}=(\tilde\pi_j)_{j=1,\ldots,s}$. The notations ``$\operatorname{post}$", ``$\operatorname{likelihood}$" and ``$\operatorname{prior}$" stand for the posterior, likelihood and prior distribution of $(\delta,\tilde\pi)$. Note that we have defined $\tilde\pi$ as an inference target in addition to the discrepancy $\delta$, in order to handle the simulation noise (as we will describe momentarily). The relationship
$$p_j(i) =  d_j(i) \tilde{p}_j(i) $$
can be used to define the posterior distribution of $\pi_j(i)$ at $p_j(i) $.

The likelihood for \eqref{Bayesian update} is straightforward to compute as
\begin{align}
\operatorname{likelihood}(d,\tilde{p},\text{data})  \propto \exp &\left( \sum_{j=1}^{s} \sum_{i=1}^{m} n_{j}(i) \log\left( d_{j}(i) \tilde{p}_j(i) \right)  + \sum_{j=1}^{s} \sum_{i=1}^{m} \tilde{n}_j(i) \log \tilde{p}_j(i)\right). \label{eq:likelihood}
 \end{align}

We now discuss the prior for \eqref{Bayesian update}. We restrict ourselves to independent priors for the discrepancy and the simulation model. The prior on the simulation model needs to exhibit the properties of a valid distribution. These properties can be enforced by conditioning an arbitrary prior distribution on a vector which takes real values in a space $\mathbb{R}^{sm}$ on the constrained region associated with Definition \ref{def:valid_distribution}. Similarly, the properties of a valid discrepancy can be enforced by conditioning an arbitrary prior distribution on the constrained region associated with Definition \ref{def:valid}. More precisely, let the logarithm of this arbitrary prior mass function for the simulation model be denoted with $f$ and the discrepancy with $g$.  Our construction leads to
\begin{equation}
\text{prior}(d,\tilde{p}) \propto \begin{cases} \exp\left(f(\tilde{p}) +  g(d) \right) &\text{ if }  \begin{cases} \tilde{p}_j(i) \geq 0, &  1\leq i \leq m,1\leq j \leq s  \\
 d_j(i) \geq 0,  &  1\leq i \leq m,1\leq j \leq s \\
\sum_{i=1}^m \tilde{p}_j(i) =1,   & 1\leq j \leq s\\
\sum_{i=1}^m \tilde{p}_j(i) d_j(i) =1, & 1\leq j \leq s
\end{cases} \\
0 &\text{otherwise}. \end{cases}  \label{eq:prior} \end{equation}

The choices of $f(\cdot)$ and $g(\cdot)$ are open to the investigator. For computational reasons that will be detailed in Section \ref{sec:optim}, we prefer $g(\cdot)$ that is concave. One widely used option that exhibits this property will be a multivariate Gaussian with a mean $\mu$ and correlation  matrix $R$ that borrows information across design points and observation points.  It is recommended that one uses a vector of $1$s as the prior mean for $\delta$ and  $(1/m)$s for $\tilde{\pi}$.   $R$ should be built with domain specific logic, e.g., similar design points and/or similar responses should have similar discrepancies. For more detailed ideas toward constructing correlation structures for responses, see \cite{ankenman2010stochastic} on the topic of stochastic kriging. In general, this approach leads to
\begin{align}
 \exp\left(f(\tilde{p}) +  g(d) \right) \propto  \exp & \left( -\lambda_{\tilde{p}} (\tilde{p} - 1/m)^\mathsf{T} R_{\tilde{p}}^{-1} (\tilde{p} - 1/m) -\lambda_d (d - 1)^\mathsf{T} R_d^{-1} (d - 1) \right), \label{eq:Gaussian}
\end{align}
where the $\tilde{p}$ and the $d$ are understood to be vectorizations of the probability masses represented by themselves, and $\lambda$s are positive constants that scale the correlation matrices $R$s.

Note that, in the settings where simulation is cheap and $\tilde\pi$ is estimated with negligible error, one can drop the parameter $\tilde p$ in the likelihood \eqref{Bayesian update} and correspondingly the second terms in the likelihood \eqref{eq:likelihood} and the prior \eqref{eq:prior}.

\section{Optimization-based Procedure for Bayesian Inference} \label{sec:optim}
This section presents our computation procedure to make conclusions about $\pi_j$ based on \eqref{Bayesian update}. In particular, we propose an optimization-based approach.  There are two reasons for considering this inference package in place of the more traditional Markov Chain Monte Carlo. First, a typical decision-making in simulation analysis often boils down to the estimation of expectation-type quantities of interest evaluated at $\pi_j$. The optimization we study will provide efficiently computable bounds on these expectations. Second, because of the constrained structure of the prior distribution \eqref{eq:prior}, standard sampling-based Bayesian computation tools are deemed to be inefficient, and optimization serves as a competitive alternative.

To elaborate the second rationale, note that common solution mechanisms in Bayesian inference consist of drawing samples from the posterior of the parameters of interest. However, because the posterior is often not a standard distribution like Normal (and that there is an unknown proportionality constant), direct Monte Carlo sampling is not possible.  Sophisticated Markov chain Monte Carlo samplers were designed explicitly for this purpose \cite[Chapters 11 and 12]{gelman2014bayesian}.  Popular samplers include the classic Metropolis Hastings algorithm with a symmetric proposal \cite[pp 278-280]{gelman2014bayesian}, and other useful methods such as Hamilton Monte Carlo \citep{duane1987hybrid} and slice sampling \citep{neal2003slice}.  The latter two methods are specifically designed to alleviate the problems faced by classical samplers.  But there are still many practical issues for these new samplers regarding their execution and choices of parameters in constrained and high dimensional spaces, which is the setting we encounter in the posterior induced from \eqref{eq:prior} (probabilistically constrained and with dimension $sm$).  See, for example, \cite{betancourt2017conceptual} for an intuitive history and theoretical summary of these conclusions.  It should be acknowledged that theoretical results do not always reveal these practical issues; see, for example, the positive results from \cite{dyer1991random}. However, numerical tests in \cite{plumlee2016learning} demonstrate these issues in a closely related setting.

In the following subsections, we will present our optimization formulation, the statistical guarantees, and discussion on computational tractability.  The summaries of the sections are:  1) We use an uncertainty set in place of a typical Bayesian integration; 2) The method is guaranteed to produce tight bounds that will contain the truth with the typical desired confidence; and 3) Given we simulate enough, the optimization problem can be reformulated into a convex problem.

\subsection{Optimization Formulation}\label{sec:formulation}
Suppose we are interested in estimating quantities of interest in the form $E[z(Y_j)]$ where $Y_j\sim\pi_j$ and $z:\mathcal Y\to\mathbb R$  is some function. We can write this in terms of $\delta$ and $\tilde\pi$ as $$\zeta(\delta,\tilde\pi)= \sum_{i=1}^m z(i) \pi_j(i) =\sum_{i=1}^mz(i)\delta_j(i)\tilde\pi_j(i).$$ Our procedure consists of solving the optimization pairs
\begin{equation}\begin{array}{ll}
\max \text{ or } \min_{d,\tilde{p}} & \zeta(d,\tilde{p}) =  \sum_{i=1}^m z(i) d_j(i) \tilde{p}_j(i), \\
\text{subject to}& \operatorname{post}(d,\tilde{p},\text{data}) \geq c \end{array} \label{obj:optim}
\end{equation}
where $c$ is chosen such that
\begin{equation}
  c = \exp \left(-\frac{1}{2} \Phi^{-1}(q)^2 + \max_{d,\tilde{p}} \log \text{post}(d,\tilde{p},\text{data})\right),\label{choice}
\end{equation}
and $ \Phi^{-1}(q)$ is the standard Normal quantile at level $q$. The optimal values of these optimization problems form an approximate confidence interval for $E[z(Y_j)]$ at a confidence level in the frequentist sense, as we will describe in Section \ref{sec:theo}.

Optimization problems  \eqref{obj:optim} can be motivated from a robust optimization viewpoint. This  literature uses deterministic sets, the so-called ambiguity or uncertainty sets, to represent the probabilistic uncertainty in the parameters  (e.g., \cite{ben2002robust,ben2009robust,bertsimas2011theory}). Typically, these sets are chosen as prediction sets that contains the truth with a prescribed confidence. The optimal values of the resulting robust optimizations then bound the true quantity of interest with at least the same confidence level. This approach has been applied in many contexts, such as approximating chance-constrained programs (e.g., \cite{ben2002robust}, Chapter 2) and performance measures driven by complex stochastic models (e.g., \cite{bandi2012tractable,bandi2014robust}). Here, we consider using a prediction set given by a posterior high probability region
\begin{equation}
\mathcal{U}(c) = \left\{d,\tilde{p} \left| \operatorname{post}(d,\tilde{p},\text{data}) \geq c \right.\right\} \label{uncertainty set}
\end{equation}
as the set of points $(d,\tilde p)$ with posterior probability higher than level $c$. From the view of robust optimization, if $c$ is chosen such that $\mathcal U(c)$ contains $1-\alpha$ posterior content of $(\delta,\tilde\pi)$, the optimal values of \eqref{obj:optim} will form an interval covering at least $1-\alpha$ posterior content of $\zeta(\delta,\tilde\pi)$.

% is larger than the choice that gives exactly $1-\alpha$ posterior content of $(\delta,\tilde\pi)$, meaning that the uncertainty set $\mathcal U(c)$ is smaller than the corresponding set.
Instead of looking for an exact $(1-\alpha)$-content prediction set, we choose our $c$ based on asymptotic theory that guarantees an asymptotically exact coverage of the true value of $E[z(Y_j)]$, which in general can be different from the choice discussed above. Our result that justifies this approach has a similar spirit to some recent studies in calibrating uncertainty sets in distributionally robust optimization,  a setting in which the uncertainty is on the underlying distribution in a stochastic problem, via asymptotic analysis based on empirical likelihood (\cite{lam2016recovering,duchi2016statistics,blanchet2016sample,lam2017empirical}) and Bayesian methods \citep{gupta2015near}. Despite these connections, to our best knowledge, there has been no direct attempt in using robust optimization as a principled Bayesian computation tool.
% , e.g., \cite{ben2009robust}, P.33 discussion point B, that stipulates the unnecessity for uncertainty sets to contain most of the probabilistic content. This observation is further studied

Our procedure essentially recovers the quantiles of the quantity of interest directly from the posterior distribution, which is the aforementioned goal of our Bayesian analysis and is conventionally obtained from sampling (e.g., Markov chain Monte Carlo). To intuitively explain the connection, consider the case when the posterior is normalized such that
\[\int_{d,\tilde p} \text{post}(d,\tilde p,\text{data}) \mathrm{d}  d \mathrm{d}\tilde p = 1.\]
The described quantile is defined as
\begin{equation}
\min\left\{a\in\mathbb R: \int_{\zeta(d,\tilde p)\leq a}  \text{post}(d,\tilde p,\text{data}) \mathrm{d}  d \mathrm{d}\tilde p \geq1-\alpha\right\}\label{quantile def}
\end{equation}
Assume that for every $a$ in consideration, there exists $(d,\tilde p)$ such that $\zeta(d,\tilde p)=a$. Then \eqref{quantile def} is equal to
\begin{equation}\begin{array}{ll}
\min_{a}&  \left\{\begin{array}{ll}\max_{d,\tilde{p}}  & \zeta(d,\tilde{p}) \\
\text{subject to}&\zeta(d,\tilde{p}) \leq a\end{array}\right\}\\
\text{subject to}&  \int_{\zeta(d,\tilde p)\leq a}  \text{post}(d,\tilde p,\text{data}) \mathrm{d}  d \mathrm{d}\tilde p  \geq1-\alpha
\end{array} \label{quantile reformulation}
\end{equation}
Denote $\mathcal U_a=\left\{d,\tilde{p} \left| \zeta(d,\tilde{p}) \leq a \right. \right\}$. We can further rewrite \eqref{quantile reformulation} as an optimization over the collection of sets in the form $\mathcal U_a$, given by
\begin{equation}\begin{array}{ll}
\min_{\mathcal U_a}&  \left\{\begin{array}{ll}\max_{d,\tilde{p}}  & \zeta(d,\tilde{p}) \\
\text{subject to}&(d,\tilde{p}) \in\mathcal U_a\end{array}\right\}\\
\text{subject to}&  \int_{\mathcal U_a} \text{post}(d,\tilde p,\text{data}) \mathrm{d}  d \mathrm{d}\tilde p  \geq1-\alpha
\end{array} \label{MC_sampler}
\end{equation}
Suppose there exists an optimal solution $\mathcal U^*$ to the outer optimization in \eqref{MC_sampler}. We conclude that the $q$ quantile of $\zeta(d,\tilde p)$ under $\text{post}(\cdot,\text{data})$ is equal to $\max_{(d,\tilde{p})\in \mathcal U^*} \zeta(d,\tilde{p}) $. Our chosen uncertainty set $\mathcal U(c)$ turns out to bear a similar performance in bounding the quantity of interest as the set $\mathcal U^*$, despite the potential vast difference in their geometries.

% Consider the presence of a sampler that can find $K$ samples, labeled by $(d^{1} ,\tilde{p}^{1}),\ldots, (d^{K} ,\tilde{p}^{K})$ from the posterior distribution which we use to approximate the posterior  distribution. The quantile can be computed in terms of a set parameterized by an $a$ labeled $\mathcal U_a=\left\{d,\tilde{p} \left| \zeta(d,\tilde{p}) \leq a \right. \right\}$,  i.e.
% \begin{equation}\begin{array}{ll}
% \min_{a}&  \left\{\begin{array}{ll}\max_{d,\tilde{p}}  & \zeta(d,\tilde{p}) , \\
% \text{subject to}&(d,\tilde{p})  \in\mathcal U_a\end{array}\right\}\\
% & \left[\# k \text{ such that } \zeta(d^{k} ,\tilde{p}^{k}) \leq a\right] \geq  K q
% \end{array} \label{MC_sampler}
% \end{equation}
% which gives the exact quantile for a distribution approximated by sampling.

\begin{figure}[htb]
{
\centering
\includegraphics[width=5in]{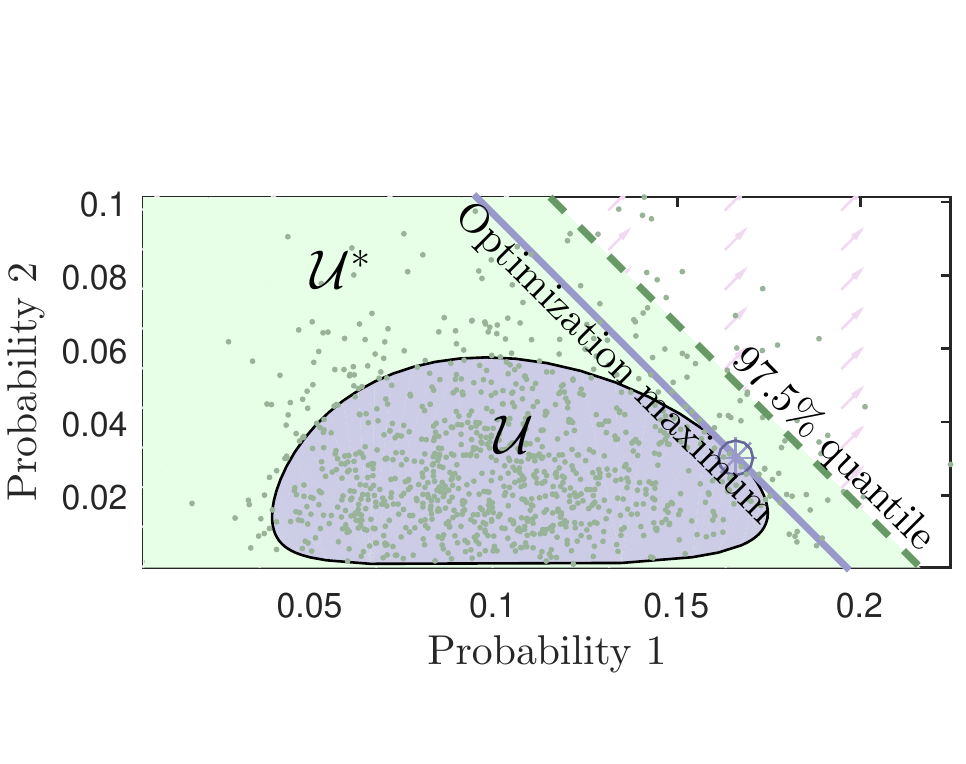}
\caption{Graphical description of the differences between optimization- and sampling-based approaches where the objective is to bound the sum of the probabilities.  The $1000$ dots are samples from the posterior.  The upper limit of the $97.5 \%$ quantile of the sum is indicated by the dashed line where $\mathcal{U}^*$ is the solution to the outer optimization in (\ref{MC_sampler}) given that these samples are the entirety of the posterior distribution.    The region labeled $\mathcal{U}$ is the projection of $\mathcal{U}$ onto this two-dimensional plane and the maximum of the optimization is determined by the solid line with $c=2$.   \label{fig:graphical_ROvSAMPLE}}
}
\end{figure}

To illustrate graphically the difference between sampling quantiles and the optimization approach, suppose we are trying to find the $97.5\%$ confidence level upper bounds for the sum of two probabilities in our system. Figure \ref{fig:graphical_ROvSAMPLE} illustrates this with samples imposed on top of the projection of the uncertainty set in \eqref{uncertainty set}, and it shows the similarity of the bounds provided by the two approaches.  Clearly, $\mathcal{U}(c)$ is much smaller compared to $\mathcal{U}^*$, yet the resulting bounds are quite similar. The next subsection investigates the properties of $\mathcal U(c)$ and explains such a phenomenon.
\subsection{Theoretical Guarantees} \label{sec:theo}
We first study the asymptotic behavior of the optimal values in \eqref{obj:optim}.  We will consider a more general setting in which the objective function is $ \sum_{j=1}^s\sum_{i=1}^m z_j(i) p_j(i)$ for some functions $z_j:\mathcal Y\to\mathbb R$, i.e., a linear combination of individual expectations at $x_j$. Evidently, $z_j(i)=0$ for all but one $j$ will reduce to the setting in \eqref{obj:optim}. For ease of exposition, define
\[Z_j \stackrel{\text{dist}}{=}z_j(Y_j),\]
where $Y_j\sim\pi_j$. Let $\mathcal{U}_n (c) $ be defined as in \eqref{uncertainty set}, with the subscript $n=\sum_{j=1}^sn_j$ indicating the total number of observed responses on the real system. Similarly, let $\text{post}_n(d,\tilde p)$ represent the posterior function when the data contains $n$ observations.

We have the following result (which is shown as Lemma \ref{lem:op_consistency} in the appendix):
\begin{theorem} \label{thm:op_consistency}
Suppose that $\tilde{\pi}_j(i)>0$ and $\pi_j(i)>0$ for all $i = 1,\ldots,m$ and $j = 1,\ldots,s$.  For each observation, the design point is an independent random variable with sample space $\{x_1,\ldots,x_m\}$ and respective positive probabilities $\xi_1,\ldots, \xi_m$.

Let $\text{post}_n^* = \max_{d,\tilde{p}} \text{post}_n (d,\tilde{p})$ and $\hat{\pi}_i^n (i) = n_j(i)/n_j .$  Then for all $\ell >0$,
\begin{equation}
\lim_{n \rightarrow \infty} \sqrt{n}  \left(\max_{(d,\tilde p)\in\mathcal{U}_n(\text{post}_n^*-\ell^2/2)} \sum_{j=1}^s\sum_{i=1}^m z_j(i) d_j(i)\tilde p_j(i) - \sum_{j=1}^s \sum_{i=1}^m z_j(i) \hat{\pi}_j^n(i)    \right)  =    \ell \sqrt{\sum_{j=1}^s \xi_j^{-1} \mathbb{V} Z_j} \label{eq:optim1}
\end{equation}
and
\begin{equation}
\lim_{n \rightarrow \infty} \sqrt{n} \left( \min_{(d,\tilde p)\in\mathcal{U}_n(\text{post}_n^*-\ell^2/2)} \sum_{j=1}^s\sum_{i=1}^m z_j(i)d_j(i)\tilde p_j(i) - \sum_{j=1}^s \sum_{i=1}^m z_j(i) \hat{\pi}_j^n(i)    \right)  =    -\ell \sqrt{\sum_{j=1}^s \xi_j^{-1} \mathbb{V} Z_j} \label{eq:optim2}
\end{equation}

almost surely, where $\mathbb{V} $ represents the variance.
\end{theorem}

An immediate observation of Theorem \ref{thm:op_consistency} is that the simulation replication size $\tilde n_j$ plays no role in the asymptotic behavior of the optimization output as $n$ gets large. Thus, with enough real data, the accuracy of the simulation runs is inconsequential, as the values of the real data dominate the results.
% Thus, for this section, we view $\tilde{n}$ as fixed without any loss of generality or fidelity in the results.
The same observation also holds for the prior choices made for  $f(\cdot)$ and $g(\cdot)$. This asymptotic independence of the prior resembles the classic Bernstein-von Mises theorem. In summary, our optimization approach generates bounds in tight asymptotic agreement with those obtained from the typical data-only inference approaches.
% Nor is there any dependence on the accuracy of the simulation and the simulation size $\tilde n_j$.
% Secondly, Theorem \ref{thm:op_consistency} implies that as the amount of collected data from the real system gets large
% , also plays no role in the inference

It is known that not every posterior distribution is guaranteed to have appropriate consistency properties; see the works of \cite{freedman1963asymptotic} and \cite{diaconis1986consistency}.    Bayesian credible sets resembling the form of \eqref{uncertainty set} are not guaranteed to produce rational inference; for more information on the general properties of Bayesian credible sets, see   \cite{cox1993analysis} or \cite{szabo2015frequentist}.  In particular, two complications arise in proving Theorem \ref{thm:op_consistency}.  First, the measure associated with the likelihood function only concentrates on a lower dimensional manifold (dimension $sm-s$) of the parameter space (dimension $sm$). This issue is by-and-large a technical one and is addressed in Lemmas \ref{lemma:str_consis} and \ref{lemma:weak_consis} proved in the appendix.    Second, the optimization problem requires a particular shape of the uncertainty set to yield the desired asymptotic properties. As a main observation, the uncertainty set $\mathcal U_n(\text{post}_n^*-\ell^2/2)$ can be shown to asymptotically become an ellipsoid, and optimization problem \eqref{obj:optim} therefore reduces to a quadratic program with an elliptical constraint, which can be analyzed and elicits the convergence behavior in Theorem \ref{thm:op_consistency}.
% toThe second issue is relatively novel and the proof is thus sketched below.

% \proof{Sketch of Proof for Theorem \ref{thm:op_consistency}}
% Theorem \ref{thm:op_consistency} is a less general version of Lemma \ref{lem:op_consistency} which is listed and proved in the Appendix.  The key idea there is to observe that, as $n$ gets large, the left hand side of  (\ref{eq:optim1}) or (\ref{eq:optim2})  is close to the quadratically constrained program
% \begin{align}
% \max_{t \in \mathbb{R}^{mq}} &\quad \sum_{j=1}^s \sum_{i=1}^m z(i) t_j(i) -  \sum_{j=1}^s \sum_{i=1}^m z(i) \hat{\pi}_j^n(i), \nonumber\\
% \text{subject to} & \quad\sum_{i=1}^m  t_j(i) = 1, \quad j = 1, \ldots, s, \label{eq:optimization_proof1} \\
% & \quad \frac{1}{2} \sum_{j=1}^s \sum_{i=1}^m \frac{\xi_j}{\pi_j(i)}  (\hat{\pi}_j^n(i) - t_j(i))^2 \leq \frac{\ell^2}{2}. \nonumber
% \end{align}
% Here, the values of $t_j(i)$ represents the deviations of $d_j(i) \tilde{p}_j(i)$ from $\hat{\pi}_j^n(i)$ inflated by a factor of $\sqrt{n}$ due to the central limit theorem.  Using Lagrange multipliers,
% \[t_j(i) = \hat{\pi}_j^n(i)  + \nu_{m+1} \frac{\pi_j(i)}{\xi_j}(z(i)-\nu_j),\]
% for some reals $\nu_1,\ldots,\nu_m$ and $\nu_{m+1}>0$.  The first constraint of (\ref{eq:optimization_proof1}) implies $\nu_j = \mathbb{E} Z_j$ for $i = 1,\ldots,m$.  The last constraint of (\ref{eq:optimization_proof1}) implies
% \[\nu_{m+1} = \ell \left(\sum_{i=1}^m\xi_j^{-1} \mathbb{V} Z_j \right)^{-1/2}.\]
% Plugging these values into the optimand of (\ref{eq:optimization_proof1}) yields the result.
% \Halmos
% \endproof

There are several implications of Theorem \ref{thm:op_consistency}.  The first is that both the upper and lower limits provided by the optimization converge to the true value almost surely as the data gets large, as described below:
\begin{corollary}
Under the same assumptions in Theorem \ref{thm:op_consistency}, we for all $\ell>0$,
$$\max_{(d,\tilde p)\in\mathcal{U}_n(\text{post}_n^*-\ell)} \sum_{j=1}^s\sum_{i=1}^m z_j(i)d_j(i)\tilde p_j(i)\to\sum_{j=1}^s\sum_{i=1}^m z_j(i)\pi_j(i)$$
and
$$\min_{(d,\tilde p)\in\mathcal{U}_n(\text{post}_n^*-\ell)} \sum_{j=1}^s\sum_{i=1}^m z_j(i)d_j(i)\tilde p_j(i)\to\sum_{j=1}^s\sum_{i=1}^m z_j(i)\pi_j(i)$$
almost surely as $n\to\infty$.\label{consistency}
\end{corollary}

Corollary \ref{consistency} shows that  with enough data the proposed posterior estimate is a good representation of the truth. It is a basic property that is in line with Bayesian consistency results studied traditionally by statisticians \citep{schwartz1965bayes}.

Furthermore, Theorem \ref{thm:op_consistency} also implies that, as $n$ gets large,
\begin{equation}
\max_{(d,\tilde p)\in\mathcal{U}_n(\text{post}_n^*-\ell^2/2)} \sum_{j=1}^s\sum_{i=1}^m z_j(i)d_j(i)\tilde p_j(i) \approx\sum_{j=1}^s \sum_{i=1}^m z_j(i) \hat{\pi}_j^n(i)    + \ell \sqrt{\frac{\sum_{j=1}^s \xi_j^{-1} \mathbb{V} Z_j}{n}} \label{asymptotic max}
\end{equation}
and
\begin{equation}\min_{(d,\tilde p)\in\mathcal{U}_n(\text{post}_n^*-\ell^2/2)} \sum_{j=1}^s\sum_{i=1}^m z_j(i)d_j(i)\tilde p_j(i) \approx\sum_{j=1}^s \sum_{i=1}^m z_j(i) \hat{\pi}_j^n(i)    - \ell \sqrt{\frac{\sum_{j=1}^s \xi_j^{-1} \mathbb{V} Z_j}{n}} \label{asymptotic min}
\end{equation}
Note that the left hand sides of \eqref{asymptotic max} and \eqref{asymptotic min} are precisely the classical confidence bounds on $\sum_{j=1}^s\sum_{i=1}^m z_j(i) \pi_j(i)$ generated from the central limit theorem with $n_j \approx \xi_j n$. This hints at a proper coverage in large samples at the level $1-\alpha$. In fact, we have the following result:
\begin{corollary}
Under the same assumptions in Theorem \ref{thm:op_consistency}, we have
$$\mathbb P\left(\max_{(d,\tilde p)\in\mathcal{U}_n(\text{post}_n^*-\ell^2/2)} \sum_{j=1}^s\sum_{i=1}^m z_j(i)d_j(i)\tilde p_j(i)\geq\sum_{j=1}^s\sum_{i=1}^m z_j(i)\pi_j(i) \right)\to \Phi(\ell)$$
and
$$\mathbb P\left(\min_{(d,\tilde p)\in\mathcal{U}_n(\text{post}_n^*-\ell^2/2)} \sum_{j=1}^s\sum_{i=1}^m z_j(i)d_j(i)\tilde p_j(i)\leq\sum_{j=1}^s\sum_{i=1}^m z_j(i)\pi_j(i)\right)\to \Phi(\ell)$$
as $n\to\infty$, where $\mathbb P$ denotes the probability generated from a data set of size $n$.\label{CI}
\end{corollary}

The above results reveal that the proposed inference differs from purely empirical estimates only when data is sparsely collected.   If data from the real system is abundant, our simulation models $\tilde{\pi}_1(\cdot),\ldots,\tilde{\pi}_s(\cdot)$ will have very little impact on our resulting conclusions.  In a sense, the Bayesian approach automatically balances the influences from the empirical data versus the simulation model. Complement to our asymptotic result in this section, our numerical examples in Section \ref{sec:illustration} will demonstrate that the difference in inference between our approach and one that ignores the simulation model can be sizable in sparse data environments.

We conclude this section by presenting a result on the consistency of a ``ranking and selection" task:
\begin{corollary} \label{corr:op_consistency}
Suppose the conditions and definitions of Theorem \ref{thm:op_consistency}.  For all $\ell>0$, if $j$ and $k$ are such that  $\mathbb{E} Z_j >  \mathbb{E} Z_k $, then
\[\lim_{n \rightarrow \infty} \mathbb P \left( \min_{(\tilde{p},d) \in \mathcal{U}(\text{post}_n^*-\ell)} \sum_{i=1}^m z_j(i)d_j(i)\tilde p_j(i) < \max_{(\tilde{p},d) \in\mathcal{U}(\text{post}_n^*-\ell)} \sum_{i=1}^m z_k(i)d_k(i)\tilde p_k(i) \right)  = 0 .\]\label{rs}
\end{corollary}
Corollary \ref{rs} implies that the intervals for the quantities of interest at different design points do not overlap as the data gets large, if their values are truly different.  Thus, in practice, a user who notes that the two intervals generated from the optimization problems do not overlap can reasonably conclude there is a difference between the two values.

\subsection{Solvability of the optimization} \label{sec:optim}
This subsection discusses the tractability of the imposed optimization problems in Section \ref{sec:formulation}. We focus on the convexity of the problems which, in contrast to the previous section,  will  depend on the replication size from the simulation model.

  To begin, we write optimization (\ref{obj:optim}) in full (focusing only on the minimization problem) as
\begin{equation}\begin{array}{ll}
\min_{d,\tilde{p}} & \zeta(d,\tilde{p}) =  \sum_{i=1}^m z_j(i) d_j(i) \tilde{p}_j(i), \\
\text{subject to}& f(\tilde{p}) + g(d)  + \sum_{j=1}^{s} \sum_{i=1}^{m} n_{j} (i) \log( d_j(i) \tilde{p}_j(i)) + \sum_{j=1}^{s} \sum_{i=1}^{m} \tilde{n}_{j}(i) \log \tilde{p}_j(i) \geq \log(c)\\
&\tilde{p}_j(i) \geq 0 , d_j(i) \geq 0, \text{ for all } i,j \\
&\sum_{i=1}^m \tilde{p}_j(i) =1, \sum_{i=1}^m \tilde{p}_j(i) \delta_j(i) =1, \text{ for all } j
\end{array} \label{obj:optim1}
\end{equation}
This formulation is generally non-convex because of the non-convex objective function and the non-convex constraint $\sum_{i=1}^m \tilde{p}_j(i) \delta_j(i) =1$, regardless of the sample sizes and the priors $f(\cdot)$ and $g(\cdot)$. However, noting that the program is individually convex in $d$ and $\tilde p$, one approach is to use alternating minimization, by sequentially optimizing $d$ fixing $\tilde p$ and $\tilde p$ fixing $d$ until no improvement is detected. Though it does not guarantee a global solution, this approach has been shown to be effective for certain chance-constrained programs (see, e.g., \cite{chen2010cvar,zymler2013distributionally,jiang2016data}).
% Consider a hypothetical there was no Monte Carlo uncertainty and the simulation model density could be exactly evaluated.  decision variable $\tilde{p}$ can be replaced with the fixed $\tilde{\pi}$. Under this situation,  problem (\ref{obj:optim}) can be rewritten as

On the other hand, supposing that there is no simulation error in estimating $\tilde\pi$, then the prior  on $\tilde p$ and the associated calculations can be removed, resulting in
\begin{equation}\begin{array}{ll}
\max _{d} & \sum_{i=1}^m z(i)\tilde{\pi}_j(i)d_j(i)   , \\
\text{subject to}& g(d)  + \sum_{j=1}^{s} \sum_{i=1}^{m} n_{j}(i) \log d_{j}(i) \geq \log(c) \\
&d_j(i) \geq 0, \text{ for all } i,j \\
 &\sum_{i=1}^m \tilde{\pi}_j(i) d_j(i) =1, \text{ for all } j
\end{array}  \label{optimization no MC}
\end{equation}
If in addition the function $g(\cdot)$ is a concave function, then \eqref{optimization no MC} is a convex optimization problem. We summarize this as:
\begin{proposition}
Problem (\ref{optimization no MC}) is a convex program if $g(\cdot)$ is a concave function on $\mathbb{R}^{sm}$.
\end{proposition}
% The set of feasible solutions of

Recalling our discussion in Section \ref{sec:learn}, one example of a concave $g(\cdot)$ corresponds to the multivariate Gaussian prior of (\ref{eq:Gaussian}).

Formulation \eqref{optimization no MC} can be reasonably used in situations where simulation replications are abundant, so that the simulation outputs are very close to $\tilde\pi_j$. Our next result shows that, in the case that $\tilde n_j$ is sufficiently large and $g(\cdot)$ satisfies a slightly stronger condition, using \eqref{obj:optim1} also leads to a convex problem. To prepare for this result, we rewrite the decision variables in \eqref{obj:optim1} to get
\begin{equation}\begin{array}{ll}
\max _{p,\tilde{p}} & \sum_{i=1}^j z(i) p_j(i), \\
\text{subject to}& f(\tilde{p}) + h(p,\tilde{p})  + \sum_{j=1}^{s} \sum_{i=1}^{m} n_{j} (i) \log p_j(i) + \sum_{j=1}^{s} \sum_{i=1}^{m} \tilde{n}_{j}(i) \log \tilde{p}_j(i) \geq \log(c)\\
&\tilde{p}_j(i) \geq 0 , p_j(i) \geq 0, \text{ for all } i,j \\
&\sum_{i=1}^m \tilde{p}_j(i) =1, \sum_{i=1}^m p_j(i) =1, \text{ for all } j
\end{array} \label{optimization reformulation}
\end{equation}
where $h(p,\tilde{p}) = g(p_j/\tilde{p}_j)$ with the operation $p_j/\tilde p_j$ defined component-wise. We recall the definition that a function $r(\cdot)$ is strongly concave if for all $a$ and $b$ and $0 \leq \lambda \leq 1$,
\[r(\lambda a+ (1-\lambda) b ) \geq \lambda r(a)+ (1-\lambda) r(b) + \beta \lambda (1-\lambda) \|a-b\|^2,  \]
where $\|a-b\|$ is the Euclidean norm and  $\beta$ is some positive constant \cite[pp 60] {nesterov2003introductory}. Our result is:

\begin{theorem} \label{thm:convex}
Assume that $g(\cdot)$ is strongly concave and differentiable on $\mathbb R_+^{sm}$ and the derivative is bounded on all compact sets in $\mathbb R_+^{sm}$, $f(\cdot)$ is bounded from above, and $\tilde\pi_j(i)>0$ for all $i,j$.

Let $\mathcal{U}_{\tilde{n}} (c_{\tilde{n}})$ be the set of feasible solutions for (\ref{optimization reformulation}) where
\[\log c_{\tilde{n}} = -\frac{\ell^2}{2} + \max_{p,\tilde{p}}\left\{  f(\tilde{p}) + h(p,\tilde{p}) + \sum_{j=1}^{s} \sum_{i=1}^{m} n_j(i) \log p_j(i) + \sum_{j=1}^{s} \sum_{i=1}^{m} \tilde{n}_{j}(i) \log \tilde{p}_j(i)\right\},\]
for some constant $\ell > 0$.  Then as $\tilde{n} \rightarrow \infty$, $\mathbb{P} \left( \mathcal{U}_{\tilde{n}} (c_{\tilde{n}}) \text{ is convex}\right) \rightarrow 1. $
\end{theorem}

Thus, given access to sufficient computing resources and properly choosing $g(\cdot)$, one can use a convex optimization solver to carry out our proposed approach, no matter how few or many data were collected from the real system. Note that this observation holds even when $f(\cdot)$ is not concave. Theorem \ref{thm:convex} hinges on a joint convexity argument with respect to $(d,\tilde{p})$ in the asymptotic regime as $\tilde n$ grows but $n$ is fixed.

\section{Numerical Illustrations } \label{sec:illustration}
We demonstrate our approach with two real-data examples. First is a proof-of-concept investigation in modeling a call center. Second is on the support of staffing decision in a  manufacturing production line discussed in the introduction of this article.

\subsection*{Call center example}\label{sec:call center}
Consider the call center data originally analyzed in \cite{brown2005statistical}.  This dataset is associated with a call center where a customer calls in and is placed a queue until one of $x$ servers is available. From these data, the sample mean of the waiting time (from entry to service for a customer) from 9:00 to 10:00 am is calculated.   In this narrow time period, the arrival rate and service rate, which is time inhomogenous according to \cite{brown2005statistical}, should be approximately homogenous.  Here, we also account for the number of servers operating in the system at any given time, which appears to differ between days (see the appendix for details). To our reading,  this subset of the dataset was by-and-large ignored in \cite{brown2005statistical}'s original analysis.

Our model for this call center will be an $x$-server first-come-first-serve queue.  Following practice, both the interarrival and service times are modeled as exponentially distributed.  After a warm-up period, the sample average of the waiting time is measured over the course of a one-hour window.    This, in principle, agrees with \cite{brown2005statistical}.    In the spirit of ad-hoc calibration, two additional features were added:  (i) the arrival rate is randomly generated each day from a log-normal distribution with associated mean $1.8$ and variance $0.4$, and (ii) a customer will abandon the queue if the waiting time is longer than an exponential random variable with mean $5$.  Adding both of these features resulted in a simulation model that was closer to the observed data.

The response is discretized into the four categories $<1$, $1-2$, $2-3$, and $> 3$ minutes ($m=4$) and we study $5-9$ servers ($s = 5$).    No data from the real system is observed at either $5$ or $9$ servers. The simulation model was evaluated 250 times at each design point.
\begin{figure}[htb]
{
\centering
\includegraphics[width=\textwidth]{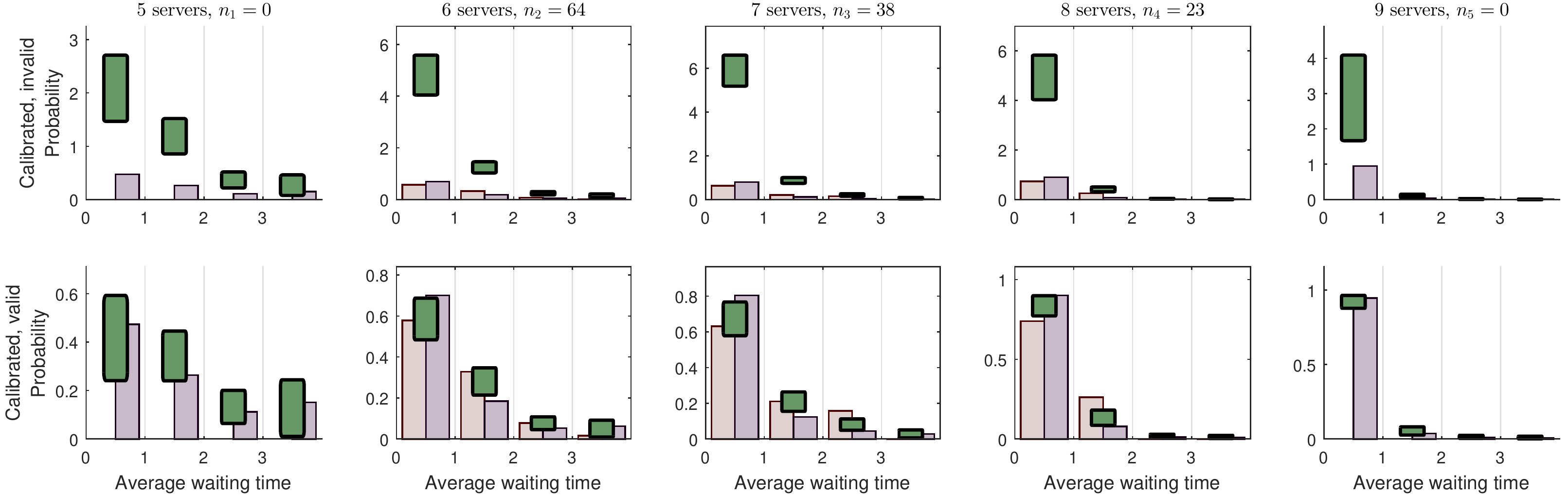}
\caption{The predictive intervals for $\pi$ used for the call center study described in Section \ref{sec:illustration}.  The subplots from left to right show the results with 5 to 9 servers, respectively.  The solid line is the observed frequency from the data when it is available at $6-8$ server levels.  The dashed line is the observed frequency from the simulation model.  The rectangles represent intervals predictions from either using the typical intervals from traditional sampling (left rectangles) or the optimization (right rectangles). } \label{fig:orig_data_calibration}
}
\end{figure}

Figure \ref{fig:orig_data_calibration} shows the intervals implied by the proposed posterior distribution using a sampling-based approach and our optimization approach using \eqref{obj:optim} and \eqref{choice}.  The functions $f$ and $g$ were of Gaussian form (\ref{eq:Gaussian}), with correlation $0.75^{|x_i-x_j| } \cdot 0.75^{|k-l| }$  between the $i$th and $j$th staffing levels and the $k$th and $l$th outputs with $\lambda_d = 1/4$ and $\lambda_p = 1/100$.  The key is the ability to answer question such as: how likely is it that the average waiting time when there are $9$ servers is between $2$ and $3$ minutes?  There is no data, but the simulation model combined with the observed responses and our prior information gives us an estimate of somewhere less than $15 \%$.  This accounts for the discrepancy that we observed based on the recorded responses at $6-8$ servers as well as the potential Monte Carlo error from running a finite number of simulations.  Overall, the ranges at other staffing levels appear to agree with both the data and the simulation outputs.    We are not confident, for example, that staffing $5$ servers will produce the same results as the simulation model, which has average waiting times over $3$ minutes about $15 \%$ of the time.  Based on the recorded responses, this could be  $30 \%$, but it could also be as low as about $1\%$.

The above discussion offers some preliminary validity check on the practical implementation. Next we illustrate the theoretical discussions in Section \ref{sec:theo}. For this purpose, consider an example where the true model is specified by us, some data is generated from this model, and an inexact simulation model is specified.

We use the same simulation model. In the dataset we found that waiting times are under-estimated by the simulation model when many servers are present.  To replicate this, we add onto our ``true" model an event (according to a Poisson process, average 5 min between events) in which if there are $5$ idle servers, all idle servers will take a break (average 30 minutes, exponentially distributed) and if there are more than $7$ idle servers, these additional severs will stop servicing for the remainder of the hour.  This will naturally inflate the waiting times.  While not exactly mimicking the real system, this reflects the general phenomena that all operators may not be working at all times in a call center.  Thus even though $8$ servers may be ``working'', because of miscellaneous personnel reasons, the queue behaves differently than the simulation model.  All other features of the true model are exactly the same as the simulation model, including the arrival rates and departure rates.

In this numerical experiment there are either $5$, $10$, $20$, $200$ or $2000$ total observations.  Two observation schemes are examined:  in the first, each observation comes from one of the staffing levels 6, 7 and 8 with equal probabilities of $1/3$;  in the second, each observation comes from one of the staffing levels 5, 6, 7, 8 and 9 with equal probabilities of $1/5$.

\begin{figure}[htb]
{
\centering
\includegraphics[width=\textwidth]{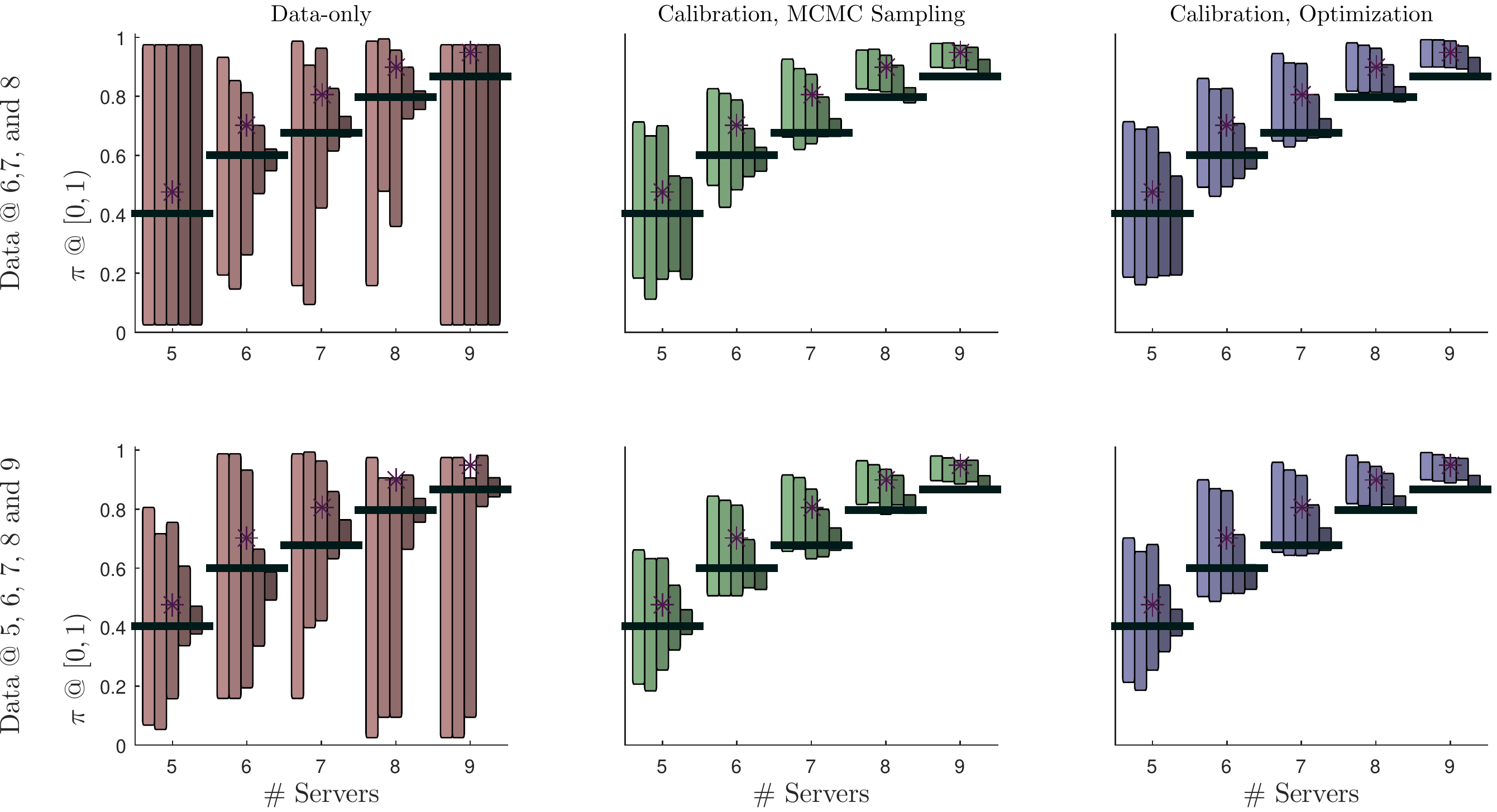}
\caption{The bounds produced in the call center example in Section \ref{sec:illustration}.   The rectangles in the six panels are decided by the data-only approach (left), the sampling-based calibration approach with $2.5\%$ and $97.5 \%$ quantiles (middle), and  the proposed optimization-based calibration approach.   The set of $5$ rectangles for each number of servers  represent  $5$, $10$, $20$, $200$ and $2000$ observations, respectively.  The long horizontal line represents the true value.    The top set of panels refers to the case where data is observed only at  6, 7, and 8 and the bottom set of panels refers to the case where data is observed at 5, 6, 7, 8 and 9.  \label{fig:ROvSAMPLE_p}}
}
\end{figure}

 Figure \ref{fig:ROvSAMPLE_p} shows the prediction of the probability the average waiting time will be less than $1$ minute.   We compare to a data-only approach which consists of bounds based on the classic confidence interval with binomial responses (either less than one minute or not).  All approaches behave similarly when the amount of data is large, agreeing with Theorem \ref{thm:op_consistency}.   But there are differences in the data-poor performances.  Consider the first observation scheme, where no data is collected at $5$ and $9$ servers.  The proposed approach correctly predicts the chance of a short average waiting time with $9$ servers to be large, while the data-only approach does not have access to the simulation model and thus predicts the chance of a short average waiting time with $9$ servers to be possibly small (the prediction covers all possibilities).   Moreover, the data-only approach can be quite poor when only a few data points exist. The conclusions reached from using the posterior with either traditional sampling or our optimization approach are comparable in the large data cases, but do differ in the small data cases.  The computational speed of the optimization approach was orders of magnitude smaller for this example compared to the sampling.

\subsection*{Manufacturing line example}
This subsection uses our calibration framework to assist a decision process for staffing a real production line.   A major manufacturer of automobiles has two parallel production lines, labeled box and closure, that suffer from frequent failures.   These failures are predominately handled by a group of workers trained to quickly identify and resolve small issues.  Due to the time needed to traverse the line combined with the relative frequency of failures, four workers are currently staffed in this support position.   The manufacturer is interested in the impact of this staffing level on the throughput of the line, measured in units per hour.    The lines' behavior are classified into thirteen categories from 46 to 74 in $2$ units per hour increments.

\begin{figure}[t]
{
\centering
\includegraphics[width=6.5in]{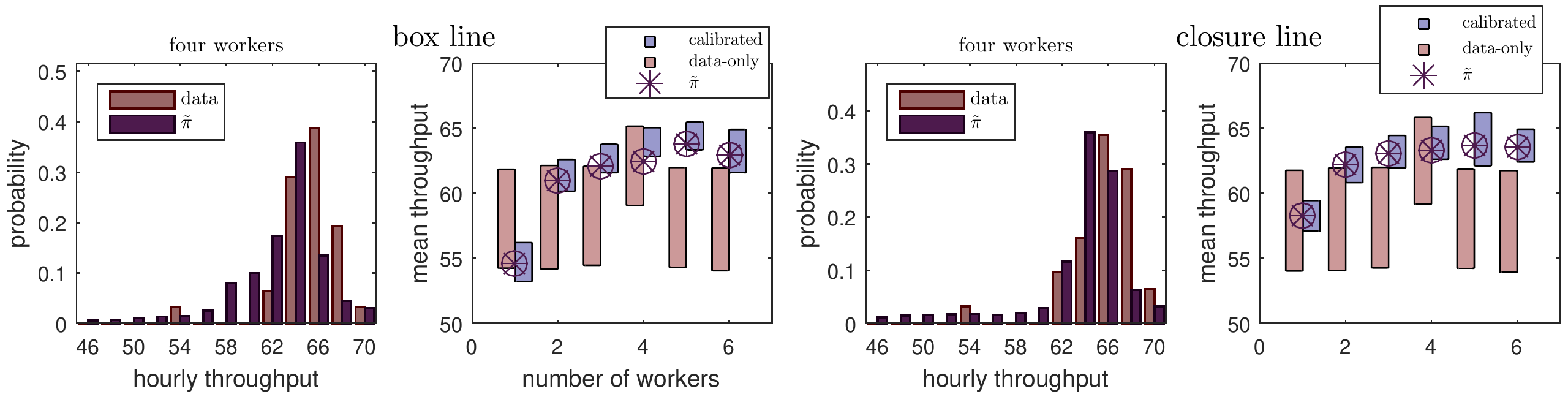}
\caption{Illustration for the manufacturing case study presented in Section \ref{sec:illustration}.  Subplots far left and middle right are for the box and closure lines comparing the model histogram and the frequency histogram from the data.  Subplots middle left and far right show the predictive intervals from the case study presented in Section \ref{sec:illustration} for the box and closure lines, respectively.  \label{fig:illustration}}
}
\end{figure}

The two lines have different criteria for ill-performance.   The box line will starve the next line if the throughput drops below 60.  The closure line will starve the next line if the throughput drops below 56.  The goal is thus to ensure that the chance of starving the next line remains near the current level when there are $4$ workers.  Since experiments on the real system would be extremely costly and potentially dangerous, an outside company was hired to design a discrete-event simulation model to investigate potential staffing reconfigurations for this group of workers.  Additionally, a two-person internal team was tasked with refining and adjusting the simulation model via ad-hoc calibration, including detailed input analysis that broke down failure rates by stations along the line.  Despite these extensive and costly efforts, the simulation model did not perfectly agree with the data collected in the current four worker configuration (see Figure \ref{fig:illustration}) due to several assumptions made along the model development process. These included typical input assumptions like independent and exponentially distributed inter-failure times as well as more complicated structural assumptions such as workers returning to their station in between maintenance calls.  Roughly $75$ realizations from the simulation were completed at each design point, as decided was sufficient toward the end of the project.

Knowing this simulation model is not perfect, what would be a reasonable estimate for the mean throughput of the line at each staffing level from $1$ worker to $6$ workers?  If we can define the priors $f$ and $g$, then this becomes an answerable question using the method described in this article.  Like the previous example, the functions $f$ and $g$ were of Gaussian form, of (\ref{eq:Gaussian}), with correlation $0.75^{|x_i-x_j| } \cdot 0.9^{|k-l| }$  between the $i$th and $j$th staffing levels and the $k$th and $l$th possible throughputs with $\lambda_d = 1/4$ and $\lambda_p = 1/100$.     This agreed, as best as possible, with the expectations of the builders of the simulation model, who think that there is a large correlation across outputs (i.e. a similar likelihood ratio at places close in the sample space) and smaller amounts of correlation across the inputs (i.e. the builders are unsure of the behavior of likelihood ratio across the input variables, but generally anticipate it is close for similar staffing levels).

Figure \ref{fig:illustration} displays the lower and upper bounds on the probabilities of low production for each line constructed from our method.  As we move away from our observations at a staffing level of $4$, the predictive bounds on the mean throughput get larger and become closer to the simulation model.  This expansion of predictive intervals and  the regression to  the simulation model mimic what is seen in calibration of deterministic models \citep{kennedy2001bayesian} and stochastic kriging \citep{ankenman2010stochastic}.   Around $4$ workers, the assumption is that that there is some correlation between staffing levels that decays as we move away from a staffing level of $4$, thus expanding our predictive intervals.

 In terms of comparison to a data-only alternative, there is clearly no ability to distinguish between different staffing levels using data alone.  In terms of an answer to the fundamental question posed by the manufacturer, a few things can be gleaned from these bounds.  For example, it becomes clear from this analysis that staffing a single worker would in high likelihood starve the next lines, which is the core problem the manufacturer would like to avoid.  The ultimate decision from the manufacturer was to do a field study of the three worker staffing level.  This was based on both feasibility assurance provided by the simulation model and the potential benefit of redeploying a worker into a different position.

\section*{Acknowledgements}
We thank Ilan Guedj for the data organization and for Avi Mandelbaum for continuing to place the data on the website \href{http://ie.technion.ac.il/serveng/}{http://ie.technion.ac.il/serveng/}. Additional thanks are due to the
Tauber Institute for Global Operations  at the University of Michigan, Anthony Sciuto, Anusuya Ramdass and Brian Talbot. We also gratefully acknowledge support from the the National Science Foundation under grants CMMI-1542020, CMMI-1523453 and CAREER CMMI-1653339.

\bibliographystyle{informs2014}
\bibliography{references}

\section*{Appendix: More information on data use}
Dates used
\begin{description}
\item[January] 4, 5, 6, 7, 14, 15, 19, 20, 21, 26, 27, 28
\item[February] 1, 2, 3,  8, 15, 16, 17, 18, 19, 22, 23, 24, 25
\item[March] 1, 4, 8, 9, 10, 15, 17, 18, 19, 22, 23, 25, 29, 30
\item[April] 2, 5, 8, 12, 13, 14, 15, 26, 27, 28, 29, 30
\item[May] 3, 4, 5, 6, 7, 10, 11, 12, 13, 14, 18, 19, 20, 24, 25, 26, 27
\item[June] 1, 2, 3, 4, 7, 8, 9, 10, 11, 14, 15, 16, 17, 21, 22, 23, 24, 28, 29, 30
\item[July] 2, 6, 7, 8, 12, 13, 14, 15, 19, 20, 21, 26, 27, 28, 29
\item[August] 4, 5, 11, 16, 17, 18, 19, 23, 24, 25, 26, 30, 31
\item[September] 1, 2, 3, 6, 7, 8, 13, 14, 15, 16, 21, 22, 23, 27, 30
\item[October] 4, 5, 6, 7, 11, 12, 13, 14, 18, 19, 20, 21, 25, 26, 27, 28
\item[November] 3, 4, 5, 8, 9, 10, 17, 18, 25, 29, 30
\item[December] 1, 2, 6, 7, 8, 9, 13, 14, 16, 17, 20, 21, 22, 23
\end{description}
We ignore data that ended with `IN' and `TT' because they arrive to a different queue with a different set of servers.  The number of servers was calculated by keeping track of how many users were being served at a time.  The floor of the median of this number over the one hour window when the queue has at least two persons created the value for the number of servers.

\section*{Appendix: Proofs}
Here we demonstrate two auxiliary lemmas and prove the results presented in this paper.

\begin{lemma} \label{lemma:str_consis}
Suppose $N_j$, $j = 1,\ldots,s$, is a realization of a multinomial random variable with $n$ trials and strictly positive probabilities $\xi_1,\ldots, \xi_s$.  Suppose for each $j$,  $N_j(1),\ldots,N_j(m)$  is a realization of a multinomial random variable with $N_j$ trials and strictly positive probabilities $\pi_j(1),\ldots, \pi_j(m)$. Let
$\hat{\pi}_j^n (i) = N_j(i)/N_j$. Let
\[\mathcal{K}_n = \left\{ \kappa_1 \in \mathbb{R}^{m-1},\ldots, \kappa_s \in \mathbb{R}^{m-1} \left| \hat \pi_j^n (i) +  \frac{1}{\sqrt{n}} \kappa_j(i)\geq 0,\quad  \sum_{i=1}^{m-1} \kappa_j(i) \leq 0\right.\right\},\]
\[F_n(\kappa) =\sum_{j=1}^s N_j(m)\log \left\{1- \sum_{i=1}^{m-1} \hat \pi_j^n (i) +\frac{1}{\sqrt{n}} \kappa_j(i) \right\}+\sum_{i=1}^{m-1}  N_j(i) \log \left\{ \hat \pi_j^n(i) +\frac{1}{\sqrt{n}}  \kappa_j(i)  \right\} \]
and $R_n(\kappa) = h\left( \hat{\pi}^n +n^{-1/2} \kappa \right),$ where $h$ is a continuous function that is bounded from above and finite.  Then for all $w>0$, $F_n(\kappa)+R_n(\kappa)-F_n(0)-R_n(0)$ converges uniformly on \[\mathcal{K}^w = \left\{ \kappa_1 \in \mathbb{R}^{m-1},\ldots, \kappa_s \in \mathbb{R}^{m-1} \left|  -w \leq \kappa_j(i)  \leq w\right.\right\}\] to
\[ f(\kappa) = -\frac{1}{2}\sum_{i=1}^m \frac{\xi_j (1-\sum_{j=1 }^{m-1}\kappa_j(i))^2}{\pi_j(m)}  -\frac{1}{2}\sum_{j=1 }^{m-1} \frac{\xi_j \kappa_j(i)^2}{\pi_j(i)}   .\]
\end{lemma}

\proof{Proof of Lemma \ref{lemma:str_consis}.}
For sufficiently large $n$, $\hat{\pi}_j^n(i) >0$ and thus
\[F_n(\kappa) =F_n(0)+W_n(\kappa)\]
where
\[W_n(\kappa)  =\sum_{j=1}^s N_j(m) \log\left( 1 -\frac{N_j}{\sqrt{n} N_j(m) } \sum_{i=1}^{m-1} \kappa_j(i)\right)  +\sum_{i =1}^{m-1}N_j(i)    \log\left(1+  \frac{N_j}{\sqrt{n}N_j(i)} \kappa_j(i) \right).\]

Also, because $R_n(\cdot)$ is continuous and bounded on a compact set, then uniform convergence over $\mathcal{K}^w$ holds with
\[R_n(\kappa)- R_n(0) \rightarrow 0.\]
Finally, using Taylor expansion and that \[\frac{N_j^2}{n N_j(i)  } \rightarrow \frac{\xi_j}{\pi_j(i)}\] almost surely,
\[W_n(\kappa)  \rightarrow  -\frac{1}{2}\sum_{j=1}^s \frac{\xi_j (1-\sum_{i=1 }^{m-1}\kappa_j(i))^2}{\pi_j(m)}  -\frac{1}{2}\sum_{i=1 }^{m-1} \frac{\xi_j \kappa_j(i)^2}{\pi_j(i)}  ,\]
uniformly over $\mathcal{K}^w$.
\Halmos
\endproof

\begin{lemma} \label{lem:op_consistency}
Suppose $N_j$, $j = 1,\ldots,s$, is a realization of a multinomial random variable with $n$ trials and strictly positive probabilities $\xi_1,\ldots, \xi_s$.  Suppose for each $j$,  $N_j(1),\ldots,N_j(m)$  is a realization of a multinomial random variable with $N_j$ trials and strictly positive probabilities $\pi_j(1),\ldots, \pi_j(m)$.

Suppose $Z_j$ is a random variable with sample space $\{z_j(1),\ldots, z_j(m)\}$ with associated probabilities $\pi_j(1),\ldots,\pi_j(m)$.   For all $\ell > 0$, let
\[\mathcal{U}_n^{\ell} = \left\{\kappa \in \mathcal{K}_n \left| R_n(\kappa) +F_n(\kappa)+ \ell  \geq \max_{\kappa \in \mathcal{K}_n} F_n(\kappa) +R_n(\kappa) \right.\right\}.\]
Then
\begin{equation}
\sum_{j=1}^s z_j(m) + \lim_{n \rightarrow \infty}\min_{\kappa \in \mathcal{U}_n^{\ell}} \sum_{j=1}^s\sum_{i=1}^{m-1} (z_j(i)-z_j(m)) \kappa_j (i) \rightarrow    \sqrt{2 \ell\sum_{j=1}^s \frac{\text{var}(Z_j)}{\xi_j} } \text{ almost surely}.\label{eq:op2}
\end{equation}
\end{lemma}
\proof{Proof of Lemma \ref{lem:op_consistency}.}
Let
\[\mathcal{U}_n^{\ell}(0) = \left\{\kappa \in \mathcal{K}_n \left| R_n(\kappa) +F_n(\kappa) + \ell  \geq F_n(0) +R_n(0)  \right.\right\},\]
which is the uncertainty set if the maximizer on the posterior  corresponds to $\hat{\pi}^n$.  From the definition of $\mathcal{U}_n^\ell$, $\mathcal{U}_n^\ell \subset \mathcal{U}_n^\ell(0)$ for all $\ell$ and this version of the uncertainty set is larger than the original.  Thus, the left hand side of (\ref{eq:op2}) is bounded from above by
\[\sum_{j=1}^sz_j(m)  + \lim_{n \rightarrow \infty}\max_{\kappa\in \mathcal{U}_n^{\ell}(0)} \sum_{j=1}^s\sum_{i=1}^{m-1} (z_j(i)-z_j(m)) \kappa_j (i).\]
Let $\ell' > \ell+\varepsilon$ for some $\varepsilon > 0$. The left hand side of (\ref{eq:op2})  is then bounded from above by
\begin{equation}
\sum_{j=1}^sz_j(m)+ \max_{\kappa \in U^{\ell'} }   \sum_{j=1}^s\sum_{i=1}^{m-1} (z_j(i)-z_j(m)) \kappa_j(i)  + \lim_{n \rightarrow \infty} \sup_{\kappa \in U^{\ell'} \setminus \mathcal{U}_n^{\ell} (0)}    \left|\sum_{j=1}^s\sum_{i=1}^{m-1} (z_j(i)-z_j(m)) \kappa_j(i) \right|.  \nonumber
\end{equation}
where
\[U^\ell = \left\{ \kappa \in \mathcal{K}^w \left| f(\kappa) +\ell  \geq  f(0)  \right.\right\}.\]
and $f$ is defined in Lemma \ref{lemma:str_consis}. From Lemma \ref{lemma:str_consis}, $F_n(\cdot) +R_n(\cdot)$  almost surely converges uniformly on $\mathcal{K}^w$ to $f(\cdot)$ plus a constant.  Thus $U^{\ell'}  \setminus \mathcal{U}_n^{\ell} (0) \rightarrow \emptyset$ which  yields that the left hand side of (\ref{eq:op2}) is bounded from above by
\[\sum_{j=1}^sz_j(m)+ \max_{\kappa \in U^{\ell'}}   \sum_{j=1}^s\sum_{i=1}^{m-1} (z_j(i)-z_j(m)) \kappa_j(i).\]

Also, for all $\varepsilon >0$ there is a sufficiently large $n$  such that \[\max_{\kappa \in  \mathcal{K}^w} R_n(\kappa)-R_n(0)  < \frac{\varepsilon}{2}.\]  Since $\hat{\pi}^n$ is the maximizer of $F_n$, the left hand side of (\ref{eq:op2})  is bounded from below by (for sufficiently large $n$)
\[\sum_{j=1}^sz_j(m)  + \lim_{n \rightarrow \infty} \max_{\kappa \in  \mathcal{U}_n^{\ell-\varepsilon/2}(0) }   \sum_{j=1}^s\sum_{i=1}^{m-1}  (z_j(i)-z_j(m)) \kappa_j(i). \]
Let $\ell'' >0$ be such that $\ell-\varepsilon < \ell''  < \ell-\varepsilon/2$.  Thus the left hand side of (\ref{eq:op2})  is bounded from below by
\begin{equation}
\sum_{j=1}^sz_j(m)+ \max_{\kappa \in U^{\ell''} }   \sum_{j=1}^s\sum_{i=1}^{m-1} (z_j(i)-z_j(m)) \kappa_j(i) - \lim_{n \rightarrow \infty} \sup_{\kappa \in U^{\ell''}  \setminus \mathcal{U}_n^{\ell-\varepsilon/2} (0)}    \left|\sum_{j=1}^s\sum_{i=1}^{m-1} (z_j(i)-z_j(m)) \kappa_j(i) \right|.  \nonumber
\end{equation}
From Lemma \ref{lemma:str_consis},  $U^{\ell''} \setminus \mathcal{U}_n^{\ell-\varepsilon/2} (0) \rightarrow \emptyset$, which yields that the left hand side of (\ref{eq:op2}) is bounded by below by
\[\sum_{j=1}^sz_j(m)+ \max_{\kappa \in U^{\ell''}}   \sum_{j=1}^s\sum_{i=1}^{m-1} (z_j(i)-z_j(m)) \kappa_j(i).\]

Choose $w$ large enough such that $\frac{1}{2} \sum_{j=1}^s \sum_{j=1}^m \frac{\xi_j}{\pi_j(i)} \kappa_j(i)^2 \leq \ell$ everywhere in $\mathcal{K}^w$.
We now solve $\sum_{j=1}^sz_j(m) +\max_{\kappa \in U^{\ell} }   \sum_{j=1}^s\sum_{i=1}^{m-1} (z_j(i)-z_j(m)) \kappa_j(i)$,  or equivalently,
\begin{align}
\max_{\kappa \in \mathbb{R}^{sm}} &\quad \sum_{j=1}^s \sum_{i=1}^{m} z_j(i) \kappa_j(i), \nonumber\\
\text{subject to} & \quad\sum_{j=1}^m \kappa_j(i) = 0, \quad j = 1, \ldots, m, \label{eq:optimization_proof} \\
& \quad \frac{1}{2} \sum_{j=1}^s \sum_{j=1}^m \frac{\xi_j}{\pi_j(i)} \kappa_j(i)^2 \leq \ell \nonumber
\end{align}
where $\kappa$s sum to $0$ as they are differences from $\hat{\pi}$.  This then gives us that
\[\kappa_j(i) = \nu \frac{\pi_j(i)}{\xi_j}(z_j(i)-\lambda_j),\]
for some $\lambda_1,\ldots,\lambda_s$ and positive $\nu$.  Looking at the first constraint of (\ref{eq:optimization_proof}),
\[\lambda_j = \frac{\sum_{i=1}^m  \pi_j(i)z_j(i)}{\sum_{i=1}^m  \pi_j(i)} = \sum_{i=1}^m  \pi_j(i)z_j(i) = \mathbb{E} Z_i.\]
Looking at the last constraint of (\ref{eq:optimization_proof}),
\[\nu^2  = \frac{ 2\ell}{ \sum_{j=1}^s \frac{1}{\xi_j}\sum_{i=1}^m \pi_j(i) \left(z_j(i)-\sum_{k=1}^m  \pi_j(k)z_j(k)\right)^2} = \frac{2 \ell}{\sum_{j=1}^s\xi_j^{-1} v_i}.\]

Thus the left hand side of (\ref{eq:op2}) is bounded from above  by
\[\sqrt{2\ell'}  \frac{ \sum_{j=1}^s \frac{1}{\xi_j}\sum_{i=1}^{m} \pi_j(i) \left(z_j(i)^2 - z_j(i) \mathbb{E}  Z_i \right)}{ \sqrt{\sum_{j=1}^s\xi_j^{-1} v_i }}= \sqrt{2 (\ell + \varepsilon) \sum_{j=1}^s\xi_j^{-1} \text{var}(Z_j) },\]
and from below by
\[\sqrt{2 (\ell - \varepsilon) \sum_{j=1}^s\xi_j^{-1}\text{var}(Z_j) }.\]
Since $\varepsilon$ is arbitrarily small, conclude the result.
\Halmos
\endproof

\proof{Proof of Corollary \ref{consistency}.}
The corollary concludes by noting that \eqref{eq:optim1} and \eqref{eq:optim2} imply
$$\max_{(d,\tilde p)\in\mathcal{U}_n(\text{post}_n^*-\ell^2/2)} \sum_{j=1}^s\sum_{i=1}^m z_j(i) d_j(i)\tilde p_j(i) - \sum_{j=1}^s \sum_{i=1}^m z_j(i) \hat{\pi}_j^n(i)\to0$$
and
$$\min_{(d,\tilde p)\in\mathcal{U}_n(\text{post}_n^*-\ell^2/2)} \sum_{j=1}^s\sum_{i=1}^m z_j(i) d_j(i)\tilde p_j(i) - \sum_{j=1}^s \sum_{i=1}^m z_j(i) \hat{\pi}_j^n(i)\to0$$
almost surely and that $\hat{\pi}_j^n(i)\to\pi_j(i)$ almost surely.
\Halmos
\endproof

\proof{Proof of Corollary \ref{CI}.}
Consider
\begin{eqnarray}
&&\mathbb P\left(\max_{(d,\tilde p)\in\mathcal{U}_n(\text{post}_n^*-\ell^2/2)} \sum_{j=1}^s\sum_{i=1}^m z_j(i)d_j(i)\tilde p_j(i)\geq\sum_{j=1}^s\sum_{i=1}^m z_j(i)\pi_j(i) \right)\notag\\
&=&\mathbb P\Bigg(\frac{\sqrt n\left(\max_{(d,\tilde p)\in\mathcal{U}_n(\text{post}_n^*-\ell^2/2)} \sum_{j=1}^s\sum_{i=1}^m z_j(i)d_j(i)\tilde p_j(i)-\sum_{j=1}^s\sum_{i=1}^m z_j(i)\hat\pi_j^n(i) \right)}{\sqrt{\sum_{j=1}^s\xi_j^{-1}\mathbb VZ_j}}{}\notag\\
&&{}\geq\frac{\sqrt n\left(\sum_{j=1}^s\sum_{i=1}^m z_j(i)\pi_j(i)-\sum_{j=1}^s\sum_{i=1}^m z_j(i)\hat\pi_j^n(i) \right)}{\sqrt{\sum_{j=1}^s\xi_j^{-1}\mathbb VZ_j}}\Bigg)\label{interim proof}
\end{eqnarray}
Standard central limit theorem arguments reveal that
$$\frac{\sqrt n\left(\sum_{j=1}^s\sum_{i=1}^m z_j(i)\pi_j(i)-\sum_{j=1}^s\sum_{i=1}^m z_j(i)\hat\pi_j^n(i) \right)}{\sqrt{\sum_{j=1}^s\xi_j^{-1}\mathbb VZ_j}}\Rightarrow  Z$$
as $n\to\infty$, where $Z$ is a standard normal variable. Using Theorem \ref{thm:op_consistency} and Slutsky's Theorem, \eqref{interim proof} converges to $\mathbb P(\ell\geq Z)=\Phi(\ell)$. Similarly,
\begin{eqnarray*}
&&\mathbb P\left(\min_{(d,\tilde p)\in\mathcal{U}_n(\text{post}_n^*-\ell^2/2)} \sum_{j=1}^s\sum_{i=1}^m z_j(i)d_j(i)\tilde p_j(i)\leq\sum_{j=1}^s\sum_{i=1}^m z_j(i)\pi_j(i)\right)\\
&=&\mathbb P\Bigg(\frac{\sqrt n\left(\min_{(d,\tilde p)\in\mathcal{U}_n(\text{post}_n^*-\ell^2/2)} \sum_{j=1}^s\sum_{i=1}^m z_j(i)d_j(i)\tilde p_j(i)-\sum_{j=1}^s\sum_{i=1}^m z_j(i)\hat\pi_j^n(i) \right)}{\sqrt{\sum_{j=1}^s\xi_j^{-1}\mathbb VZ_j}}{}\\
&&{}\leq\frac{\sqrt n\left(\sum_{j=1}^s\sum_{i=1}^m z_j(i)\pi_j(i)-\sum_{j=1}^s\sum_{i=1}^m z_j(i)\hat\pi_j^n(i) \right)}{\sqrt{\sum_{j=1}^s\xi_j^{-1}\mathbb VZ_j}}\Bigg)\\
&\to&\mathbb P(-\ell\leq Z)=\Phi(\ell)
\end{eqnarray*}
as $n\to\infty$.
\Halmos
\endproof

\proof{Proof of Corollary \ref{corr:op_consistency}.}
It follows as a direct consequence of Corollary \ref{consistency}.
\Halmos
\endproof

\begin{lemma} \label{lemma:weak_consis}
Suppose $N_1, \ldots,N_s$  is a realization of a multinomial random variable with $n$ trials and strictly positive probabilities $\xi_1,\ldots, \xi_s$.  Further suppose for each $j$,  $N_j(1),\ldots,N_j(m)$  is a realization of a multinomial random variable with $N_j$ trials and strictly positive probabilities $\pi_j(1),\ldots, \pi_j(m)$.  Let $r$ be a function that is bounded from above and finite. Let
\[Q_n (p)  =  r(p)+\sum_{j=1}^s N_j(m)\log \left\{1- \sum_{i=1}^{m-1} p_j(i) \right\}+\sum_{i=1}^{m-1}  N_j(i) \log p_j(i). \]
  Let $\ell$ be some strictly positive constant and $\mathcal{U}_n =\left\{p \left| Q_n (p) \geq -\ell + Q_n(\pi)\right. \right\} .    $  Then if $A$ does not contain $\pi$ then
\[ \mathbb{P} (A \cap \mathcal{U}_n = \emptyset)  \rightarrow 1 \text{ as } n \rightarrow \infty.\]
\end{lemma}
\proof{Proof of Lemma \ref{lemma:weak_consis}.}

\[Q_n (p) - D_n = h(p) - h(\pi) + \sum_{j=1}^s \left[ N_j(m)  \log \frac{1- \sum_{i=1}^{m-1}  p_j(i) }{1- \sum_{i=1}^{m-1}  \pi_j(i)}+ \sum_{i=1}^{m-1} N_j(i) \log \frac{p_j(i)}{\pi_j(i)} \right].\]

Let $\varepsilon > 0$ be sufficiently small such that  $A$ has no intersection with
\[V_\varepsilon = \left\{ p \left|\sum_{j=1}^s \xi_j  \pi_j(m) \log \frac{1- \sum_{i=1}^{m-1}  p_j(i) }{1- \sum_{i=1}^{m-1}  \pi_j(i)}  + \sum_{i=1}^{m-1} \xi_j \pi_j(i)  \log \frac{p_j(i)}{\pi_j(i)}    > - \varepsilon \right. \right\}.\]
Then
\[\mathbb{P} (A \cap \mathcal{U}_n = \emptyset)  \leq \mathbb{P}\left( \sup_{p \in V_\varepsilon^c} \frac{1}{n} \left(\ell+h(p) - h(\pi) +  \sum_{j=1}^s N_j(m) \log \frac{1- \sum_{i=1}^{m-1}  p_j(i) }{1- \sum_{i=1}^{m-1}  \pi_j(i)}  +  \sum_{i=1}^{m-1} N_j(i)\frac{p_j(i)}{\pi_j(i)} \right)       < -\frac{\varepsilon}{2}\right)\]
By the law of large numbers, and that $h(p) - h(\pi)$ is bounded from above, this tends to one as $n \rightarrow \infty$ .
\Halmos
\endproof

%\subsection*{Proof of Theorem \ref{thm:convex}}
\proof{Proof of Theorem \ref{thm:convex}}
We drop the $\tilde{n}$ subscripts for this proof. Let $\mathcal{P}$ be the combination of simplices
\[\mathcal{P} = \left\{p \left| p_j(i) \geq 0, \sum_{i=1}^m p_j(i) = 1 \right.\right\}.\]
Let
\[V_a =  \left\{p \in \mathcal{P} \left| \sum_{j=1}^s \sum_{i=1}^m (p_j(i)^{-1} - \tilde{\pi}_j(i)^{-1})^2 < \frac{a^2}{4} \right. \right\}.\]
Since $g(\cdot)$ has a continuous and finite gradient by assumption, let $G$ be a constant that bounds the gradient of $g$ on the set
\[\left\{ d \left| d_j(i) = \frac{p_j(i)}{\tilde{p}_j(i)}, p \in \mathcal{P}, \tilde{p} \in V_1   \right.\right\}.\]
Let $0< \varepsilon < 1/3$ be such that
\[ \varepsilon < \frac{\beta}{2 G \sqrt{s m}} . \]

We have
\[\mathbb{P} \left(U(c) \text{ not convex} \right) \leq  \mathbb{P} \left(U(c) \text{ not subset of } (p,\tilde{p}) \in \mathcal{P} \times V_\varepsilon \right) +  \mathbb{P} \left(r \text{ not concave on } (p,\tilde{p}) \in \mathcal{P} \times V_\varepsilon \right). \]
Lemma \ref{lemma:weak_consis} can be applied showing that the probability the first term goes to $0$.  The second term will now be shown to go to zero as well.  Let
\[r(p,\tilde{p}) = f(p) + h(p,\tilde{p})  + \sum_{j=1}^{s} \sum_{i=1}^{m} n_j(i) \log p_j(i)  + \sum_{j=1}^{s} \sum_{i=1}^{m} \tilde{n}_{j}(i) \log \tilde{p}_j(i).\]
By the relationship between concave function level sets and convex sets, we need only to show that the events
\begin{equation}
\sup_{p,\tilde{p} \in \mathcal{P} \times V_\varepsilon;  p',\tilde{p}' \in \mathcal{P} \times V_\varepsilon} \sum_{j=1}^s \sum_{i=1}^m  \left(\frac{\partial r (p,\tilde{p}) }{\partial p_j(i)}  - \frac{\partial r(p',\tilde{p}') }{\partial p_j(i)}\right) (p_j(i) -p_j(i)) \leq 0 \label{eq:term1}
\end{equation}
and
\begin{equation}
\sup_{p,\tilde{p} \in \mathcal{P} \times V_\varepsilon;  p',\tilde{p}' \in \mathcal{P} \times V_\varepsilon} \sum_{j=1}^s \sum_{i=1}^m  \left(\frac{\partial r(p,\tilde{p})  }{\partial \tilde{p}_j(i)}  - \frac{\partial r(p',\tilde{p}') }{\partial \tilde{p}_j(i)}\right) (\tilde{p}_j(i) -\tilde{p}_i'(j)) \leq 0,  \label{eq:term2}
\end{equation}
happen with probability tending to one.

 Take $p,\tilde{p}$ and $p',\tilde{p}'$.  Let $\gamma_{ij}$ be the gradient of $g$ with respect to $\delta_j(i)$ at $p_j(i)/\tilde{p}_j(i)$.  Let $\gamma_{ij}'$ be the gradient of $g$ with respect to $\delta_j(i)$ at $p_j'(i)/\tilde{p}_j'(i)$. Theorem 2.1.9 from \cite{nesterov2003introductory} gives that the strong convexity of $g(\cdot)$ implies
\begin{equation}
 \sum_{j=1}^s \sum_{i=1}^m \left(\gamma_{ij} - \gamma_{ij}'  \right) \left(\frac{p_j(i)}{\tilde{p}_j(i)} - \frac{p_j'(i)}{\tilde{p}_j'(i)} \right) \leq - \beta \sum_{j=1}^s \sum_{i=1}^m \left(\frac{p_j(i)}{\tilde{p}_j(i)} - \frac{p_j'(i)}{\tilde{p}_j'(i)} \right)^2 .  \label{eq:str convexity}
 \end{equation}
Then
\begin{align}
& \sum_{j=1}^s \sum_{i=1}^m \left(\frac{\partial r (p,\tilde{p}) }{\partial p_j(i)}  - \frac{\partial r  (p',\tilde{p}')  }{\partial p_j(i)}\right) (p_j(i) -p_j'(i))  \nonumber\\
=&   \sum_{j=1}^s \sum_{i=1}^m n_j(i) \left(\frac{1}{p_j(i)} - \frac{1}{p_j'(i)} \right) \left(p_j(i) - p_j'(i)\right) +\sum_{j=1}^s \sum_{i=1}^m  \frac{\gamma_{ij}p_j(i) }{\tilde{p}_j(i)} -  \frac{\gamma_{ij}p_j'(i)}{\tilde{p}_j(i)}-\frac{\gamma_{ij}' p_j(i)}{\tilde{p}_j'(i)} +  \frac{\gamma_{ij}' p_j'(i)}{\tilde{p}_j'(i)}   \nonumber\\
\leq&  \sum_{j=1}^s \sum_{i=1}^m \left(\gamma_{ij} - \gamma_{ij}'  \right) \left(\frac{p_j(i)}{\tilde{p}_j(i)} - \frac{p_j'(i)}{\tilde{p}_j(i)} \right)  + \sum_{j=1}^s \sum_{i=1}^m  \gamma_{ij}'  \left(\frac{1}{\tilde{p}_j'(i)} -  \frac{1}{\tilde{p}_j(i)} \right)  (p_j(i)-p_j'(i)) \nonumber\\
\leq&  - \beta \sum_{j=1}^s \sum_{i=1}^m \left(p_j(i) - p_j'(i) \right)^2  + 2 \varepsilon G \sqrt{ \sum_{j=1}^s \sum_{i=1}^m   (p_j(i)-p_j'(i))^2} \nonumber
\end{align}
The first inequality comes from the negativity of the first term and rearrangement of  terms.  The second inequality is from (\ref{eq:str convexity}), $\tilde{p}_j(i) < 1$ and the Cauchy-Schwarz inequality.  We know that $ \sum_{j=1}^s \sum_{i=1}^m \left(p_j(i) - p_j'(i) \right)^2 \leq s m $ since $p,p' \in \mathcal{P}$.  Thus
\[ \sup_{p,p' \in \mathcal{P}} - \beta \sum_{j=1}^s \sum_{i=1}^m \left(p_j(i) - p_j'(i) \right)^2  + 2 \varepsilon G \sqrt{ \sum_{j=1}^s \sum_{i=1}^m   (p_j(i)-p_j'(i))^2} = 0 \]
everywhere in $V_\varepsilon$, giving us (\ref{eq:term1}) happens everywhere on $\mathcal{P} \times V_\varepsilon$.

Working through the left hand side of (\ref{eq:term2}) on $\mathcal{P} \times V_\varepsilon$,
\begin{align}
\frac{1}{\tilde{n}} \sum_{j=1}^s \sum_{i=1}^m  \left(\frac{\partial r (p,\tilde{p}) }{\partial \tilde{p}_j(i)}  - \frac{\partial r(p',\tilde{p}') }{\partial \tilde{p}_j(i)}\right) \left(\tilde{p}_j(i) -\tilde{p}_i'(j)\right)   \rightarrow& \sum_{j=1}^s \sum_{i=1}^m \tilde{\pi}_j(i)  \left(\frac{1}{\tilde{p}_j(i)} - \frac{1}{\tilde{p}_j'(i)}\right) \left(\tilde{p}_j(i) - \tilde{p}_i'(j) \right)  \nonumber\\
\leq& - \left(\min_{ij} \frac{\tilde{\pi}_j(i)}{\tilde{p}_j(i) \tilde{p}_j'(i)}\right)  \sum_{j=1}^s \sum_{i=1}^m \left(\tilde{p}_j(i)  -\tilde{p}_j'(i) \right)^2    \nonumber\\
\leq & -\min_{ij} \tilde{\pi}_j(i) \left(1-\varepsilon/2\right)^2 \sum_{j=1}^s \sum_{i=1}^m \left(\tilde{p}_j(i)  -\tilde{p}_j'(i) \right)^2.     \nonumber
\end{align}
This convergence is uniform on $\mathcal{P} \times V_\varepsilon$ via the equicontinuity of the log-likelihood function with respect to $\tilde{p}$, see e.g. \cite{rubin1956uniform}.  Thus we have that (\ref{eq:term2}) happens on $\mathcal{P} \times V_\varepsilon$ with probability tending to one.
\Halmos
\endproof

\end{document}